\documentclass[journal]{IEEEtran}
\usepackage{amsmath,amsfonts}
\usepackage{algorithmic}
\usepackage{algorithm}
\makeatletter
\newcommand{\removelatexerror}{\let\@latex@error\@gobble}
\makeatother
\usepackage{array}
\usepackage[caption=false,font=footnotesize,labelfont=rm,textfont=rm]{subfig}
\usepackage{textcomp}
\usepackage{stfloats}
\usepackage{url}
\usepackage{verbatim}
\usepackage{graphicx}
\usepackage{cite}
\usepackage{flushend}
\usepackage{graphicx}
\usepackage{float}
\usepackage{subfig}
\usepackage{booktabs} 
\usepackage{multirow}
\usepackage{color}
\usepackage[colorlinks,
linkcolor=black,
anchorcolor=black,
citecolor=black]{hyperref}
\usepackage[T1]{fontenc}
\newtheorem{remark}{Remark}
\begin{document}
	
	\title{A Dynamic UAVs Cooperative Suppressive Jamming Method with Joint Task Assignment and Bandwidth Allocation}
	
	\author{
		Ruiqing Han, Tianxian Zhang*, Han Zhong and Yuanhang Wang
		% <-this % stops a space
				\thanks{This work was supported in part by the GF Science and Technology Special Innovation Zone Project, and in part by the Fundamental Research Funds of Central Universities under Grant 2672018ZYGX2018J009.}
				\thanks{R. Han, T. Zhang and H. Zhong are with the School of Information and Communication Engineering, University of Electronic Science and Technology of China, Chengdu, Sichuan, 611731, China (E-mail: txzhang@uestc.edu.cn; tianxianzhang@gmail.com).}% <-this % stops a space
				\thanks{Y. Wang is with the School of Information and Communication Engineering, Dalian University of Technology, Dalian 116024, China.}
				\thanks{\textit{(Corresponding author: Tianxian Zhang.)}}
	}
	\maketitle
	\begin{abstract}
		The low detectability and low cost of unmanned aerial vehicles (UAVs) allow them to swarm near the radar network for effective jamming. The key to jamming is the reasonable task assignment and resource allocation of UAVs. However, the existing allocation model is somewhat ideal, weakly adaptive to the dynamic environment, and rarely considers frequency matching, which cannot suppress the frequency agile radar (FAR) network effectively. To solve these problems, a dynamic UAVs cooperative suppressive jamming method with joint task assignment and bandwidth allocation is proposed. To represent the matching relationship between UAVs and FARs, a system model of task assignment and bandwidth allocation is established, the problem is formulated as a dynamic mixed integer programming (D-MIP) problem. Then, a suppressive jamming evaluation indicator is proposed, and the utility function is designed based on the Quality of Service (QoS) framework to quantify the jamming effect of UAVs. To solve the combinational optimization problem, a two-step dynamic hybrid algorithm based on Kriging model is proposed, which can obtain the task assignment and bandwidth allocation schemes of UAVs by consuming fewer computational resources \textcolor{black}{in dynamic environment.} Simulation results show that the proposed method is effective in terms of jamming performance, computational resource saving and dynamic environment adaptability.
		
	\end{abstract}
	
	\begin{IEEEkeywords}
		Unmanned aerial vehicles (UAVs), frequency agile radar (FAR) network, task assignment, bandwidth allocation
	\end{IEEEkeywords}
	%, mixed integer programming (MIP), dynamic optimization, surrogate model
	\section{INTRODUCTION}
	\IEEEPARstart{I}{n}	modern electronic warfare (EW), it is essential for jammers to disrupt the efficient use of the electromagnetic (EM) spectrum by radar\cite{de2018introduction}. The radar anti-jamming technology is increasingly sophisticated\cite{10330755}. Frequency agile radar (FAR) counteracts jammers by changing the carrier frequency over a wide bandwidth unpredictably \cite{yi2022adaptation,Axelsson2007AnalysisOR}, making it difficult to match the jamming frequency and reducing the jamming effect. In addition, by communicating information and sharing resources among multiple radars, a radar network could be constructed\cite{Charlish2020TheDF,Shi2020LowPO}, which greatly increases the difficulty of jamming. To counter the increasingly advanced radar network system, the research on multi-jammer network has been greatly developed\cite{neng1995survey,deligiannis2016power}. 
	The low detectability and low cost of unmanned aerial vehicles (UAVs) allows them to swarm close to radar network for suppressive jamming\cite{Siddiqi2022AnalysisOS,liu2022hybrid}, thus covering the high-value target for penetration. As a result, \textcolor{black}{UAVs swarm jamming is increasingly becoming a anticipated field\cite{wyh}.} However, the jamming performance of UAVs is highly dependent on the task assignment and resource allocation of UAVs. Due to the different jamming performance of UAVs on different radars, inappropriate task assignment and resource allocation schemes will significantly reduce the jamming performance of UAVs. Therefore, it is \textcolor{black}{essential} to \textcolor{black}{investigate} the appropriate task assignment and resource allocation method of UAVs.

	In general, the task assignment and resource allocation of UAVs face three main challenges: 1) The allocation model is somewhat ideal and the jamming evaluation indicator is difficult to establish; 2) Due to the low load capability of UAVs, the hardware computing performance of jammers are limited; 3) The task assignment and resource allocation schemes of UAVs are dynamic due to the fickle battlefield situations (i.e., the movement of covered high-value target). In recent years, \textcolor{black}{several scholars have investigated} the problem of jamming task assignment and resource allocation in other scenarios \cite{zhang2021joint,li2022robust,zhai2009iiga}. 
	In these studies, the jamming task assignment is mainly characterized by the selection of transmission beams, and the resource allocation is mostly the allocation of power resources. 
	The working frequency of the radar network in the above studies is fixed and does not take into account the bandwidth allocation of UAVs, as mentioned in challenge 1) where the allocation model is somewhat ideal. The task assignment and resource allocation schemes are often obtained by using evolutionary algorithms (EAs), such as genetic algorithm (GA) and particle swarm optimization (PSO)\cite{kennedy1995particle,Holland1975AdaptationIN}, which require a lot of computational resources corresponding to challenge 2). In addition, the task assignment and resource allocation schemes in the above researches usually remain unchanged, which cannot be adapted to the dynamic task assignment and resource allocation problems in challenge 3). 
	Obviously, these existing task assignment and resource allocation methods are not \textcolor{black}{applicable to the problem of} joint task assignment and bandwidth allocation considered in this article. 
	Therefore, it is essential to establish a task assignment and bandwidth allocation model of UAVs, 
	propose a reasonable jamming evaluation indicator to quantitatively evaluate the jamming effect on the FAR network, and study a dynamic method with joint task assignment and bandwidth allocation under limited computing resources to suppress FAR network effectively.

Fortunately, researches solving the mixed integer programming (MIP) problem with EAs provide valuable inspiration for the joint task assignment and bandwidth allocation problem of UAVs cooperative suppressive jamming. In fact, EAs have been widely application to MIP problems in various areas, such as massive multiple-input multiple-output systems \cite{liu2016energy}, delta-connected switched capacitors \cite{su2015pso}, and power generation expansion planning \cite{sirikum2007new}. For solving MIP problems, EAs are combined with integer-restriction-handling techniques and constraint-handling techniques to deal with discrete variables and constraints efficiently \cite{lin2001co,farmani2003self,cai2006multiobjective}.
However, when the battlefield situation is fickle and the UAV hardware has limited computational capability as mentioned in challenges 2) and 3), traditional EAs could not adapt to the changing environment once converged\cite{yang2005memory}, and the high dimensionality of the variables causes EAs to consume a lot of computational resources. Therefore, EAs cannot be used to effectively solve MIP problems in such dynamic environments with limited computational resources.

 In this paper, the joint task assignment and bandwidth allocation problem is a dynamic mixed integer programming (D-MIP) problem with limited computational resources due to the dynamic task assignment and bandwidth allocation schemes and the limited computational performance of UAVs. For dynamic optimization problem (DOP), several methods have been developed into EAs to solve DOP, such as random immigrants\cite{vavak1996comparative}, increasing diversity\cite{cobb1993genetic}, memory storage\cite{branke1999memory}, and multi-population approaches\cite{branke2000multi}. Among these methods, random immigrants and memory storage have been proven to be effective for a lot of DOPs\cite{yang2005memory}. For the case of limited computational resources, surrogate model-assisted evolutionary algorithms (SAEAs) are widely used in engineering design to reduce the computational burden of optimization process \cite{holena2012surrogate},\cite{liu2016surrogate}. The basic idea is to create an approximation function that approximates the real response, which can be quickly calculated and used to guide searches, rather than expensive evaluations. However, EAs integrating DOP methods and surrogate model to solve the D-MIP problem are rarely reported. Hence, in this paper, we attempt to combine the DOP methods and the surrogate model into EAs to propose a resource-saving dynamic allocation algorithm.

\textcolor{black}{In summary, different from our previously published work that only considered the problem of jamming task assignment of UAVs\cite{wyh}, in this article we study the joint jamming task assignment and bandwidth allocation problem of UAVs. We propose a dynamic UAVs cooperative suppressive jamming method with joint task assignment and bandwith allocation against FAR network.
In this paper, we construct a detailed system model including task assignment model and bandwidth allocation model. To suppress FAR network effectively, we present reasonable jamming effect evaluation indicator and devise appropriate quantification function. In addition, we propose a resource-saving dynamic allocation algorithm which is effective in terms of jamming performance, computational resource saving and dynamic adaptation.}
	%Specifically, a task assignment and bandwidth allocation system model of UAVs is built to describe the matching relationship between UAVs and FARs. 
	%the problem of UAVs suppressive jamming FAR network is solved, the task assignment and bandwidth allocation of UAVS are explored, the optimization model is established to quantify the suppressive jamming effect, and a two-step dynamic hybrid algorithm integrating multiple methods is proposed. 
	The main contributions of this article are summarized below.

	\begin{enumerate}
		\item[1)] \textit{Construct a detailed task assignment and bandwidth allocation system model of UAVs to describe the matching relationship between UAVs and FARs.} The task assignment matrix of UAVs is established by Boolean variable to characterize the transmitted single beam, and the bandwidth allocation matrix of UAVs is established to describe the bandwidth resource allocation. Then, the constraints of task assignment and bandwidth allocation are defined. The optimization model is established, which is formulated as a D-MIP problem with limited computational resources.
		\item[2)] \textit{Propose a suppressive jamming effect evaluation indicator and devise a utility function to quantify the suppressive jamming effect of UAVs.} To suppress FAR network effectively, a suppressive jamming effect evaluation indicator is proposed considering the bandwidth allocation of UAVs. To quantify the suppressive jamming effect of UAVs, the utility function containing a reward function and a cost function is devised based on Quality of Service (QoS) framework, and the design criterion of utility function is proposed to guide the design of the utility function. 
		\item[3)] \textit{Propose a two-step dynamic hybrid algorithm based on Kriging model to solve the task assignment and bandwidth allocation schemes of UAVs to suppress the FAR network effectively.}  The algorithm uses GA and FROFI\cite{wang2015incorporating} to efficiently handle discrete task assignment variables and continuous bandwidth allocation variables under constraints, integrates random immigrants and memory storage methods to adapt to dynamic environment, and establishes a Kriging model\cite{lophaven2002dace} to predict the utility of different task assignment schemes, which saves computational resources.
	%		Considering the large computational burden of this problem, a Kriging model\cite{lophaven2002dace} is established to predict the utility of different task assignment schemes without searching for the feasible optimal bandwidth allocation schemes, which greatly saves computing resources.
%%		The optimization problem is decomposed into two sub-problems, which are solved by outer task assignment algorithm and inner bandwidth allocation algorithm respectively. To adapt to the dynamic environment, the algorithm integrates random immigrants and memory storage methods. Considering the large computational burden of this problem, a Kriging model\cite{lophaven2002dace} is established to predict the utility of different task assignment schemes without searching for the feasible optimal bandwidth allocation schemes, which greatly saves computing resources.
	\end{enumerate}
	
	The rest of this article is organized below. The system model of UAVs is presented in Section II. The problem formulation is given, a suppressive jamming effect evaluation indicator is proposed and the utility function is designed in Section III. In Section IV, a two-step dynamic hybrid algorithm based on Kriging model is proposed to obtain the task assignment and bandwidth allocation schemes of UAVs. 
	\textcolor{black}{In Section V, extensive numerical simulations verify the effectiveness of the proposed algorithm in the aspects of jamming performance, computational resource saving and dynamic adaptability. Section VI gives a brief conclusion to this paper.}

	\section{SYSTEM MODEL}
		\begin{figure*}[t]
		\centerline{\includegraphics[width=40pc]{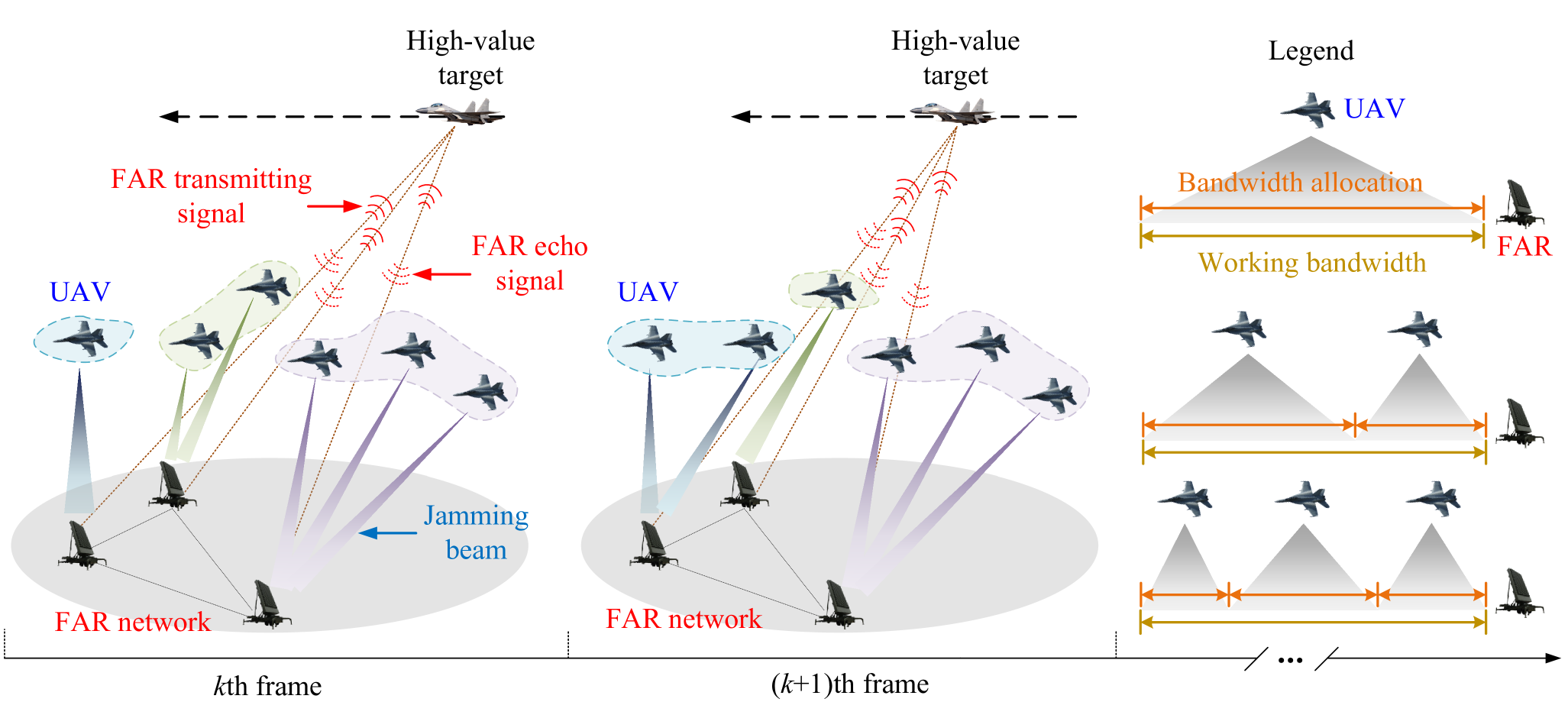}}
		\caption{Scene diagram of 6 UAVs suppressing 3 FARs to cover high-value target penetration where the legend indicates the bandwidth allocation of UAVs.}
		\label{fig111}
	\end{figure*}
	Consider the cover penetration scenario depicted in Fig. \ref{fig111}, where a high-value target carries out a penetration mission to destroy a FAR network that is constantly detecting high-value target.
	In order to cover the penetration of high-value target, a lot of UAVs suppress the FAR network, making it difficult for the FAR network to detect high-value target. 
	UAVs transmit beams to suppress the FAR network, which is defined as the task assignment of UAVs in this paper. Since the carrier frequency of FAR network is constantly changing in broadband, UAVs need to generate multiple narrow bandwidths to jointly suppress the working bandwidth of FAR network, which is defined as the bandwidth allocation of UAVs in this article. The bandwidth allocation of UAVs is shown in the legend on the right side of Fig. \ref{fig111}. If a UAV suppresses a FAR alone, the bandwidth allocation of the UAV is equal to the working bandwidth of the FAR. If several UAVs suppress a FAR together, the sum of the bandwidth allocation of UAVs is equal to the working bandwidth of the FAR. \textcolor{black}{The state of high-value target in the constant velocity (CV) model could be given by}
	\begin{equation}
		\mathbf{X}(k)=\mathbf{F}(k)\mathbf{X}(k-1)+\mathbf{\Gamma}(k) \mathbf{\nu}(k),
	\end{equation}
	where $\mathbf{\nu}(k)=\left[\nu_x,\nu_y\right]^T$ stands for the process noise vector, $\mathbf{X}(k)$ stands for the state vector of the high-value target, and its matrix forms at the $k$th suppressive jamming time step (referred to as $k$th frame in the following paper) is given by
	\begin{equation}
		\mathbf{X}(k)=\left[x^k_t,\dot{x}^k_t,y^k_t,\dot{y}^k_t\right]^T,
	\end{equation}
	\textcolor{black}{where $x_{t}^{k}$ and $y_{t}^{k}$ stand for $x$-axis and $y$-axis positions of high-value target at the $k$th frame, respectively. $\dot{x}_{t}^{k}$ and $\dot{y}_{t}^{k}$ stand for $x$-axis and $y$-axis velocities of high-value target at the $k$th frame, respectively. $\left[\cdot\right]^T$ stands for the transpose operation.} 
	$\mathbf{F}(k)$ stands for the state-transition function, which is defined as
	\begin{equation}
		\mathbf{F}(k)={
			\left[ \begin{array}{cccccc}
				1 & \Delta T & 0&0\\
				0 & 1 & 0&0\\
				0 & 0 & 1&\Delta T\\
				0&0&0&1\\
			\end{array} 
			\right ]},
	\end{equation}
	where $\Delta T$ stands for the measurement sampling interval, $\mathbf{\Gamma}(k)$ stands for the distribution matrix of process noise, which is defined as
	\begin{equation}
		\mathbf{\Gamma}(k)={
			\left[ \begin{array}{ccc}
				0.5\Delta T^2 & 0\\
				\Delta T & 0 \\
				0& 0.5\Delta T^2  \\
				0 & \Delta T  \\
			\end{array} 
			\right ]}.
	\end{equation}
	
	UAVs work in a time synchronization way and each UAV selects to suppress one or none of the FARs. 
	\textcolor{black}{The set of UAVs could be defined as $\mathcal{N}=\{1,2,\cdots,N\}$, where $N>0$ represents the number of UAVs.
	The positions of the UAVs remain unchanged and the position of the UAV $n\in\mathcal{N}$ is denoted as $(x_{j}^{n},y_{j}^{n})$, where $x_{j}^{n}$ and $y_{j}^{n}$ stand for $x$-axis and $y$-axis positions of UAV $n$ respectively.
	The set of ($M$+1) tasks could be denoted as $\mathcal{T}=\{0,1,\cdots,M\}$, where task 0 represents the UAV does not suppress any FAR, task $m$, $m=1,2,\cdots,M$ represents UAV suppresses FAR $m$.}
	It is worth noting that the system model constructed in this article requires the task assignment of UAVs to be determined before the bandwidth allocation of UAVs, in which case the UAVs can effectively suppress the FAR network to cover high-value target.
	Next, the task assignment model and bandwidth allocation model of UAVs are established respectively.
	
	\subsection{Task assignment model of UAVs}
	The task assignment matrix of UAVs at the $k$th frame could be defined by
	\begin{equation}
		\label{111111}
		\mathbf{U}_k=[\mathbf{u}_{1,k},\mathbf{u}_{2,k},\cdots,\mathbf{u}_{N,k}]^T,
	\end{equation}
	where $\mathbf{u}_{n,k}$ stands for the task assignment vector of UAV $n\in\mathcal{N}$ at the $k$th frame, and $\mathbf{u}_{n,k}$ can be respectively defined as 
	\begin{equation}
		\label{7}
		\mathbf{u}_{n,k}=[u_{n,0,k},u_{n,1,k},u_{n,2,k},\cdots,u_{n,M,k}]^T,
	\end{equation}
	where $u_{n,m,k}$ stands for the task assignment of UAV $n$ to FAR $m\in\mathcal{T}$ at the $k$th frame, which could be defined by Boolean variable as
	\begin{equation}
		\label{9}
		u_{n,m,k}=\left\{
		\begin{aligned}
			%\nonumber
			&1, &&{\rm if~UAV}~n~{\rm suppresses~ FAR}~m~{\rm at~the~}k{\rm th}\\
			& &&{\rm frame}\\
			&0, &&\rm otherwise\\
		\end{aligned},
		\right.
	\end{equation}
	as discussed above, $m=0$ means that UAV does not suppress any FAR. \textcolor{black}{Therefore, $u_{n,m,k}=1$ with $m=0$ implies that UAV $n$ does not suppress any FAR, and $u_{n,m,k}=1$ with $m\in\mathcal{T}\setminus\{0\}$ implies that UAV $n$ suppresses FAR $m\in\mathcal{T}\setminus\{0\}$, where $\mathcal{A}\setminus\mathcal{B}$ represents the exclusion of set $\mathcal{B}$ from set $\mathcal{A}$.} Since UAV $n$ selects to suppress at most one FAR, that is
	\begin{equation}
		\label{10}
		0\leq\sum_{m=0}^{M}u_{n,m,k}\leq1.
	\end{equation}
	
	\subsection{Bandwidth allocation model of UAVs}
	The bandwidth allocation matrix of UAVs at the $k$th frame could be defined by
	\begin{equation}
		\label{999}
		\mathbf{B}_k=[\mathbf{B}_{1,k},\mathbf{B}_{2,k},\cdots,\mathbf{B}_{N,k}]^T,
	\end{equation}
	where $\mathbf{B}_{n,k}$ stands for the bandwidth allocation vector of UAV $n\in\mathcal{N}$ at the $k$th frame, and $\mathbf{B}_{n,k}$ can be defined by
	\begin{equation}
		\label{8}
		\mathbf{B}_{n,k}=[B_{n,0,k},B_{n,1,k},B_{n,2,k},\cdots,B_{n,M,k}]^T,
	\end{equation}
	where $B_{n,m,k}$ stands for the bandwidth allocation of UAV $n$ to FAR $m\in\mathcal{T}$ at the $k$th frame. If UAV $n$ suppresses FAR $m\in\mathcal{T}\setminus\{0\}$, a narrow bandwidth $B_{n,m,k}>0$ is generated by UAV $n$ to suppress FAR $m$. Otherwise, $B_{n,m,k}=0$, that is
	\begin{equation}
		\label{11}
		B_{n,m,k}=\left\{
		\begin{aligned}
			%\nonumber
			&B_{n,m,k}>0,&& {\rm if}~u_{n,m,k}=1,m\in\mathcal{T}\setminus\{0\}\\
			&0,&& \rm otherwise\\
		\end{aligned}.
		\right.
	\end{equation}
	
	The bandwidth generated by a UAV is narrow, requiring multiple UAVs to jointly suppress the working bandwidth of a FAR. These UAVs that suppress the same FAR could be formed a group. The UAV group that suppresses FAR $m$ at the $k$th frame is defined as $\mathcal{C}_{m,k}$. A UAV only belongs to one group at the $k$th frame, that is $\mathcal{C}_{m,k}\cap\mathcal{C}_{\tilde{m},k}
	=\emptyset$,$\forall m,\tilde{m}\in\mathcal{T}$,$m\neq\tilde{m}$. 
	The UAVs in $\mathcal{C}_{m,k}$ have the ability to suppress FAR $m$ and reduce its detection probability, making it difficult to detect high-value target. We denote this suppressive jamming ability of UAVs in $\mathcal{C}_{m,k}$ as utility and the utility of $\mathcal{C}_{m,k}$ is $U(\mathcal{C}_{m,k})$. 
	A larger value of $U(\mathcal{C}_{m,k})$ means that a lower detection probability of FAR $m$, making it more difficult to detect high-value target. \textcolor{black}{Next, the detailed formulation derivation of $U(\mathcal{C}_{m,k})$ is presented in the next section.}

	\section{PROBLEM FORMULATION}
	%\begin{figure*}[t]
	%	\centering
	%	\subfloat[]{\includegraphics[width=2.9in]{宽带宽11.pdf}%
		%		\label{fig:a}}
	%	\hfil
	%	\subfloat[]{\includegraphics[width=2.9in]{窄带宽11.pdf}%
		%		\label{fig:b}}
	%	\caption{UAV $n$ suppresses FAR $m$ under different bandwidth allocation at the $k$th frame. (a) UAV $n$ allocates a wide bandwidth; (b) UAV $n$ allocates a narrow bandwidth.}
	%	\label{ab}
	%\end{figure*}
	
%	From the formulated problem of the task assignment and bandwidth allocation of UAVs for cooperative suppressive jamming, it is natural to devise the utility function according to the suppressive jamming reward and the system cost caused. 
	For the formulation problem of joint task assignment and bandwidth allocation of UAVs, the utility function based on suppressive jamming reward and system cost could be defined as
	\begin{equation}
		\label{3}
		U(\mathcal{C}_{m,k})=R(\mathcal{C}_{m,k})-\lambda\cdot C(\mathcal{C}_{m,k}),
	\end{equation}
	where $R(\cdot)$ stands for the reward function, $C(\cdot)$ stands for the cost function, and $0\leq\lambda\leq1$ stands for the cost factor.
	
	\begin{remark}
		\textcolor{black}{Cost factor $\lambda$ could be set flexibly according to the actual battlefield conditions.} Generally, when the detection capability of FAR network is strong and the stealth effect of UAV is poor, the probability of UAV carrying out jamming mission to be detected increases and it is more likely to be exposed, in this case, \textcolor{black}{the $\lambda$ could be set larger.} In Section V, we test the impact of different cost factors on utility and task assignment and bandwidth allocation of UAVs.
	\end{remark}

\begin{figure}[t]
	\centering
	\subfloat[]{\label{fig:a}\includegraphics[width=3in]{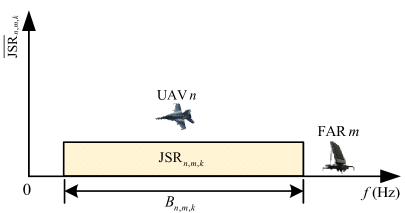}}\\
	\subfloat[]{\label{fig:b}\includegraphics[width=3in]{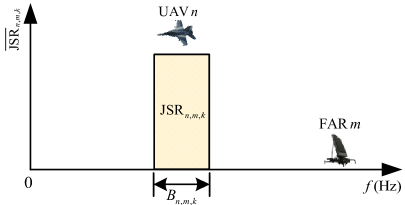}}\\	
	\caption{UAV $n$ suppresses FAR $m$ under different bandwidth allocation at the $k$th frame. (a) UAV $n$ allocates a wide bandwidth; (b) UAV $n$ allocates a narrow bandwidth.}
	\label{ab}
\end{figure}
	
	\textcolor{black}{Generally, Jamming to signal ratio (JSR) is a commonly used metric to evaluate the effect of suppressive jamming}\cite{paine2018evaluating}. 
	JSR is the ratio between the power of suppressive jamming signal and the power of high-value target echo signal. The FAR transmits signal to detect high-value target and the power of echo signal received by the FAR $m$ from the high-value target  at the $k$th frame, which could be expressed as
	\begin{equation}
		\label{13}
		P_{S}^{m,k}=\frac{P_{t}^{m}({G_{r}^{m}})^2\lambda^2\sigma}{(4\pi)^3({R_{t}^{m,k}})^4},
	\end{equation}
	where $P_{t}^{m}$ stands for the transmitting power of FAR $m$, $G_{r}^{m}$ stands for the main lobe gain of FAR $m$, $\lambda$ stands for FAR signal wavelength, $\sigma$ stands for the FAR cross section (RCS) of the high-value target, $R_{t}^{m,k}$ stands for the distance between high-value target and FAR $m$ at the $k$th frame, which could be given by
	\begin{equation}
		R_{t}^{m,k}=\sqrt{(x_{t}^{k}-x_{m})^2+(y_{t}^{k}-y_{m})^2},
	\end{equation}
	where $(x_{m},y_{m})$ is the location of FAR $m$.
	
	Under the suppressive jamming of UAV $n$, the power of suppressive jamming signal received by FAR $m$ at the $k$th frame could be expressed as 
	\begin{equation}
		\label{15}
		P_{J}^{n,m,k}=\frac{P_{j}^{n}G_{j}^{n}G_{r}^{n,m,k}\lambda^2\gamma_j}{(4\pi)^2({R_{j}^{m,n}})^2},
	\end{equation}
	where $P_{j}^{n}$ stands for the transmitting power of UAV $n$, $G_{j}^{n}$ stands for the main lobe gain of the UAV $n$, $\gamma_j$ stands for the polarization loss of the suppressive jamming signal to the FAR antenna, $R_{j}^{m,n}$ stands for the distance between UAV $n$ and FAR $m$, which could be expressed as
	\begin{equation}
		R_{j}^{n,m}=\sqrt{(x_{j}^{n}-x_{m})^2+(y_{j}^{n}-y_{m})^2},
	\end{equation}
	where $G_{r}^{n,m,k}$ is the gain of FAR $m$ for the suppressive jamming signal from UAV $n$ at the $k$th frame, which is given by\cite{graham2011communications}
	\begin{equation}
		G_{r}^{n,m,k}=\left\{
		\begin{aligned}
			%\nonumber
			&G_{r}^{m}, &&0\leq\theta^{n,m,k}\leq\frac{\theta_{0.5}}{2}\\
			&\xi\left(\frac{\theta_{0.5}}{\theta^{n,m,k}}\right)^2G_{r}^{m}, &&\frac{\theta_{0.5}}{2}\leq\theta^{n,m,k}<\frac{\pi}{2} \\
			&\xi\left(\frac{2\theta_{0.5}}{\pi}\right)^2G_{r}^{m}, &&\theta^{n,m,k}\geq\frac{\pi}{2}\\
		\end{aligned},
		\right.
	\end{equation}
	where $\theta^{n,m,k}$ stands for the angle between the maximum gain direction of UAV $n$ and FAR $m$ at the $k$th frame, $\theta_{0.5}$ stands for the main lobe width of FARs, $\xi$ stands for a constant.

	According to Eq. \eqref{13} and Eq. \eqref{15}, the JSR of UAV $n$ suppressing FAR $m$ at the $k$th frame could be expressed as
	\begin{equation}
		\label{18}
		{\rm JSR}_{n,m,k}=\frac{P_{J}^{n,m,k}}{P_{S}^{m,k}}
		=4\pi\frac{P_{j}^{n}G_{j}^{n}G_{r}^{n,m,k}\gamma_j(R_{t}^{m,k})^4}{P_{t}^{m}({G_{r}^{m}})^2\sigma({R_{j}^{m,n}})^2}.
	\end{equation}

	Based on the derivation of Eq. \eqref{18}, it can be found that the formulation of JSR does not contain frequency domain information, so JSR cannot be directly used to evaluate the suppressive jamming effect of UAVs on FAR network.
	Therefore, a novel suppressive jamming evaluation indicator needs to be proposed. Next, we design the detailed formulation of reward function and cost function.
	
	\subsection{Reward Function}

	The bandwidth allocation of UAV $n$ directly affects the suppressive jamming effect, the influence of bandwidth allocated by UAV $n$ on the suppressive jamming effect is shown in Fig. \ref{ab}. 
	The shaded area stands for ${\rm JSR}_{n,m,k}$, and the shaded areas in Fig. \ref{ab}(a) and Fig. \ref{ab}(b) are the same.
	From Fig. \ref{ab}(a), UAV $n$ allocating a wide bandwidth means that there is a high probability that UAV $n$ can suppress frequency of FAR $m$.
	The frequency of FAR $m$ jumps randomly within the working bandwidth, and the suppressive jamming probability of UAV $n$ to FAR $m$ at the $k$th frame could be expressed as
	\begin{equation}
		\label{19}
		p_{n,m,k}=\frac{B_{n,m,k}}{B_{m}},
	\end{equation}
	where $B_m$ is the working bandwidth of FAR $m$.

	Although a wide bandwidth of UAV $n$ can suppress a large range of FAR working bandwidth, the suppressive jamming energy of UAV $n$ is also dispersed over the bandwidth. We combined ${\rm JSR}_{n,m,k}$ and $B_{n,m,k}$ to define a novel suppressive jamming effect evaluation indicator as
	\begin{equation}
		\label{20}
		{\rm \overline{JSR}}_{n,m,k}=\frac{{\rm JSR}_{n,m,k}}{B_{n,m,k}},
	\end{equation}
	where ${\rm \overline{JSR}}_{n,m,k}$ stands for the novel suppressive jamming effect evaluation indicator containing frequency-domain information. A large value of ${\rm \overline{JSR}}_{n,m,k}$ implies that UAV $n$ has a significant suppressive jamming effect. When ${\rm JSR}_{n,m,k}$ is constant, larger $B_{n,m,k}$ means smaller ${\rm \overline{JSR}}_{n,m,k}$. Therefore, a wide bandwidth allocated by UAV $n$ in Fig. \ref{ab}(a) means that the suppressive jamming probability $p_{n,m,k}$ is large and the suppressive jamming effect ${\rm \overline{JSR}}_{n,m,k}$ is small. Similarly, a narrow bandwidth allocated by UAV $n$ in Fig. \ref{ab}(b) means that the suppressive jamming probability $p_{n,m,k}$ is small and the suppressive jamming effect ${\rm \overline{JSR}}_{n,m,k}$ is large.
	
	According to the above discussion, the reward function is related to the suppressive jamming effect and suppressive jamming probability. Therefore, the reward function of $\mathcal{C}_{m,k}$ could be expressed as
	\begin{equation}
		\label{1}
		R(\mathcal{C}_{m,k})=\left\{
		\begin{aligned}
			%\nonumber
			&\sum_{n=1}^{N}{f({\rm \overline{JSR}}_{n,m,k})\cdot p_{n,m,k}},&&{\rm UAV}~n\in\mathcal{C}_{m,k}\\
			&0,&&{\rm otherwise} \\
		\end{aligned},
		\right.
	\end{equation}
	where $f(\cdot)$ denotes the suppressive jamming effect function. The reward function is the sum of suppressive jamming effect into the suppressive jamming probability of each UAV in $\mathcal{C}_{m,k}$. By exploring the relationship between suppressive jamming effect function $f(\cdot)$ and ${\rm \overline{JSR}}_{n,m,k}$ sufficiently, the design of suppressive jamming effect function $f(\cdot)$ \textcolor{black}{needs to satisfy the criteria below}:
	\begin{enumerate}
		\item suppressive jamming effect function $f(\cdot)$ is a increasing function of ${\rm \overline{JSR}}_{n,m,k}$;
		\item suppressive jamming effect function $f(\cdot)$ satisfies the marginal utility theory\cite{veblen1909limitations}, i.e., the derivative of $f({\rm \overline{JSR}}_{n,m,k})$ regarding ${\rm \overline{JSR}}_{n,m,k}$ tends to be zero when ${\rm \overline{JSR}}_{n,m,k}\rightarrow0$ or ${\rm \overline{JSR}}_{n,m,k}\rightarrow+\infty$.
	\end{enumerate}
	
		\begin{figure}[t]
		\centerline{\includegraphics[width=19pc]{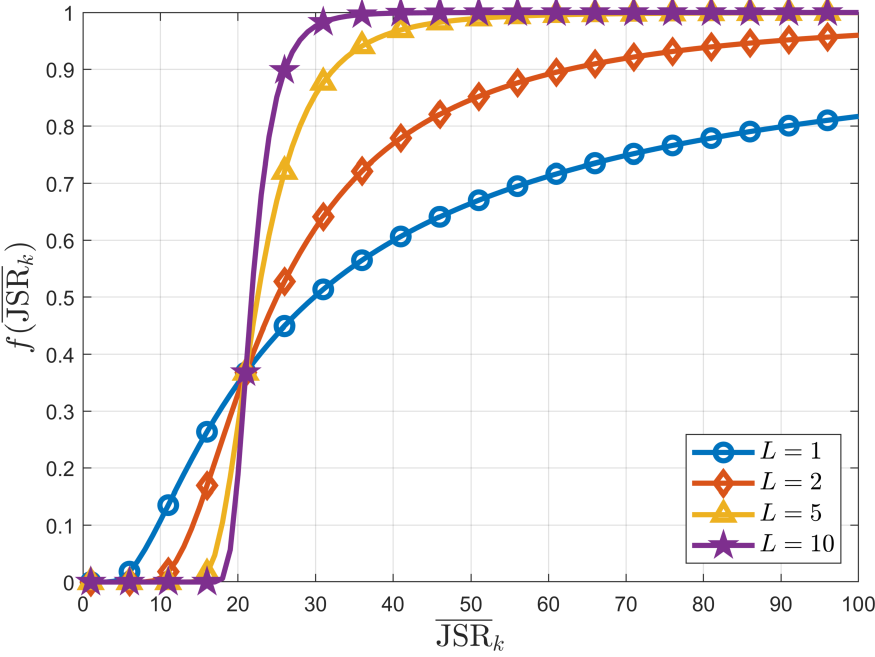}}
		\caption{Curves of suppressive jamming effect function $f({\rm \overline{JSR}}_{n,m,k})$ under different $L$.}
		\label{fig2}
	\end{figure}

	For criterion 1), as discussed above, a higher ${\rm \overline{JSR}}_{n,m,k}$ means that the suppressive jamming effect is significant. \textcolor{black}{However, the suppressive jamming effect of UAVs cannot be increased indefinitely.} When the suppressive jamming effect increases to a certain value, the detection probability of FAR network is already very low, and high-value target cannot be detected \cite{blair1998benchmark}. 
	As ${\rm \overline{JSR}}_{n,m,k}$ continues to increase, the suppressive effect will not increase significantly, that is, $f(\cdot)$ will gradually converge with the increase of ${\rm \overline{JSR}}_{n,m,k}$.
	\textcolor{black}{Therefore, it needs to satisfy criterion 2).} To realise this goal, some scholars use the sigmoid function combined with the quality of service (QoS) framework to devise the reward function of radar network task assignment to achieve efficient multi-target tracking\cite{xiong2022coalition}. 
	Based on the above inspiration, the suppressive jamming effect function $f(\cdot)$ in this work can be designed by
	\begin{equation}
		\label{12}
		\begin{aligned}
			\begin{aligned}
				f({\rm \overline{JSR}}_{n,m,k})&={\rm exp}\left[-\left(\frac{{\rm JSR}^o}{{{\rm \overline{JSR}}}_{n,m,k}}\right)^L\right]\\
				&={\rm exp}\left[-\left(\frac{{\rm JSR}^o\cdot B_{n,m,k}}{{{\rm JSR}}_{n,m,k}}\right)^L\right]
			\end{aligned}
		\end{aligned},
	\end{equation}
	where ${\rm JSR}^o$ represents the requirement of JSR for the suppressive jamming of FAR network, and $L>0$ stands for the tolerance factor. Let ${\rm JSR}^o=20$, and suppress effect function $f({\rm \overline{JSR}}_{n,m,k})$ under different $L$ is shown in Fig. \ref{fig2}. We observe that the devised suppressive jamming effect function satisfies criterion 1) and 2) and tolerance factor $L$ could determine the steepness of curves. In addition, Eq. \eqref{12} is also a decreasing function of $B_{n,m,k}$ and satisfies the marginal utility theory.

	The sum of total bandwidth of UAVs in $\mathcal{C}_{m,k}$ need to cover the working bandwidth of FAR $m$ completely, that is
	\begin{equation}
		\label{22}
		\sum_{n=1}^{N}B_{n,m,k}=B_m,~{\rm UAV}~n\in\mathcal{C}_{m,k},
	\end{equation}
	and the bandwidth allocation $B_{n,m,k}$ also need to meet some inequality constraints, that is
	\begin{equation}
		\label{23}
		0\leq B_{n,m,k}\leq B_m,~{\rm UAV}~n\in\mathcal{C}_{m,k}.
	\end{equation}
	
	\begin{figure*}[t]
		\centerline{\includegraphics[width=41pc]{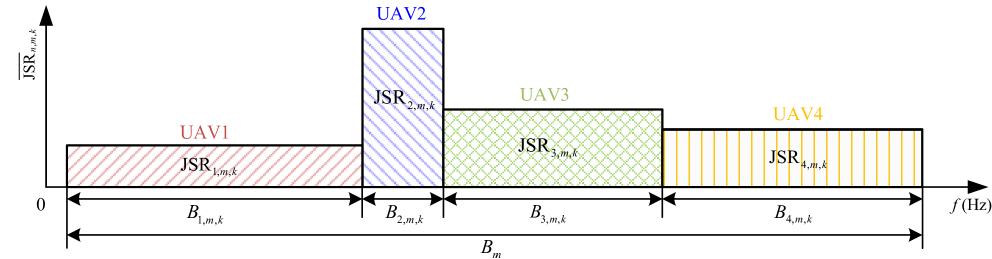}}
		\caption{An example of 4 UAVs allocating certain bandwidth to jointly suppress the FAR $m$ at the $k$th frame.}
		\label{1212}
	\end{figure*}

		\begin{figure*}[b]
		\noindent\rule{\textwidth}{1pt}
		\begin{equation}
			\label{26}
			U(\mathcal{C}_{m,k})=\left\{	
			\begin{aligned}
				%\nonumber
				&\sum_{n=1}^{N}{\exp\left[-\left(\frac{{\rm JSR}^o\cdot B_{n,m,k}}{{{\rm JSR}}_{n,m,k}}\right)^L\right]\cdot \frac{B_{n,m,k}}{B_m}}-\lambda\cdot\frac{\sum_{n=1}^{N}{u_{n,m,k}}}{N},&&{\rm UAV}~n\in\mathcal{C}_{m,k}~{\rm if}~ \mathcal{C}_{m,k}\neq\emptyset,m\in\mathcal{T}\setminus\{0\}\\
				&0,&&{\rm otherwise} \\
			\end{aligned}.
			\right.
		\end{equation}
	\end{figure*}

	For ease of understanding, a more intuitive explanation is given by taking an example of 4 UAVs allocating a certain bandwidth to suppress a FAR $m$, as shown in Fig. \ref{1212}. We find that the sum of the bandwidth allocated by 4 UAVs is equal to FAR $m$ working bandwidth $B_m$, i.e., $B_{1,m,k}+B_{2,m,k}+B_{3,m,k}+B_{4,m,k}=B_m$, satisfying Eq. \eqref{22} and Eq. \eqref{23}. UAV1 allocates a wide bandwidth $B_{1,m,k}$ so that the suppressive jamming probability $p_{1,m,k}$ is large and the ${\rm \overline{JSR}}_{1,m,k}$ is small, and UAV2 allocates a narrow bandwidth so that the ${\rm \overline{JSR}}_{2,m,k}$ is large and the suppressive jamming probability $p_{2,m,k}$ is small. 
	Multiple UAVs jointly suppress jamming FAR $m$, which not only ensures full coverage of the working bandwidth of FAR $m$, but also ensures that the jamming energy of UAVs will not be dispersed to the broadband bandwidth.

	\subsection{Cost Function}
	For UAVs in $\mathcal{C}_{m,k}$ that suppress FAR $m$, information sharing among UAVs increases the communication burden, and UAVs transmitting jamming beams increase the risk of detection by the FAR network, all of which are costs incurred when UAVs execute suppressive jamming missions. Generally, the cost function depends on the number of UAVs in $\mathcal{C}_{m,k}$, which is designed by
	\begin{equation}
		\label{2}
		C(\mathcal{C}_{m,k})=\frac{\sum_{n=1}^{N}{u_{n,m,k}}}{N},~{\rm UAV}~n\in\mathcal{C}_{m,k},
	\end{equation}
	where $\sum_{n=1}^{N}{u_{n,m,k}}$ represents the number of working UAVs in $\mathcal{C}_{m,k}$ at the $k$th frame. Eq. \eqref{2} indicates that a larger number of UAVs working in $\mathcal{C}_{m,k}$ represents a larger cost $C(\mathcal{C}_{m,k})$.

	Substituting Eq. \eqref{19}, Eq. \eqref{1}, Eq. \eqref{12} and Eq. \eqref{2} into Eq. \eqref{3}, the designed utility function of $\mathcal{C}_{m,k}$ can be rewritten in Eq. \eqref{26} as

	Considering the different threat levels of different FARs, by introducing weights $\omega_1, \omega_2,\cdots,\omega_M$, $\sum_{m=1}^{M}\omega_m=1$, the utility of all UAVs could be expressed as
	\begin{equation}
		\label{1111}
		U(\mathcal{C}_{k})=\sum_{m=1}^{M}\omega_mU(\mathcal{C}_{m,k}),
	\end{equation}
	where $\mathcal{C}_{k}$ represents the UAV swarm consisting of all UAV groups at the $k$th frame, a larger $\omega_m$ means that the FAR is more threatening, and if more UAVs choose to suppress this FAR, their suppressive jamming utility will be better.

	The problem of joint task assignment and bandwidth allocation is established as a maximization utility $U(\mathcal{C}_{k})$ problem, and constraints such as Eq. \eqref{10}, Eq. \eqref{22} and Eq. \eqref{23} are taken into account. In conclusion, the optimization model could be formulated as
	\begin{equation}
		\label{27}
		\begin{aligned}
			&\max~U(\mathcal{C}_{k})\\
			&s.t.\left\{
			\begin{aligned}
				%\nonumber
				&0\leq\sum_{m=0}^{M}u_{n,m,k}\leq1,&&\forall~n\in\mathcal{N}\\
				&\sum_{n=1}^{N}B_{n,m,k}=B_m,&&\forall ~m\in\mathcal{T}\setminus\{0\}\\
				&0\leq B_{n,m,k}\leq B_m,&&\forall ~m\in\mathcal{T}\setminus\{0\}
			\end{aligned}
			\right.
		\end{aligned}.
	\end{equation}

	Note that this optimization problem in Eq. \eqref{27} is a challenging combinatorial optimization problem. 
	%As is mentioned in Section I, the task assignment and bandwidth allocation of UAVs is both a MIP problem and a DOP, and the limitation of actual computing resources should also be considered. EAs provide a good tool for solving MIP problems, random immigrants and memory schemes are also applied to DOP, and surrogate models are widely used to reduce the computational burden. 
	The task assignment of UAVs is a discrete variable processing problem, and the bandwidth allocation of UAVs is a continuous variable processing problem with some constraints. Meanwhile, the task assignment and bandwidth allocation schemes of UAVs are dynamic and the hardware computing capacity of UAVs are limited. Therefore, the mathematical nature of this problem could be formulated as a dynamic mixed integer programming (D-MIP) problem with limited computational resources.
	
\begin{figure}[t]
	\centerline{\includegraphics[width=20pc]{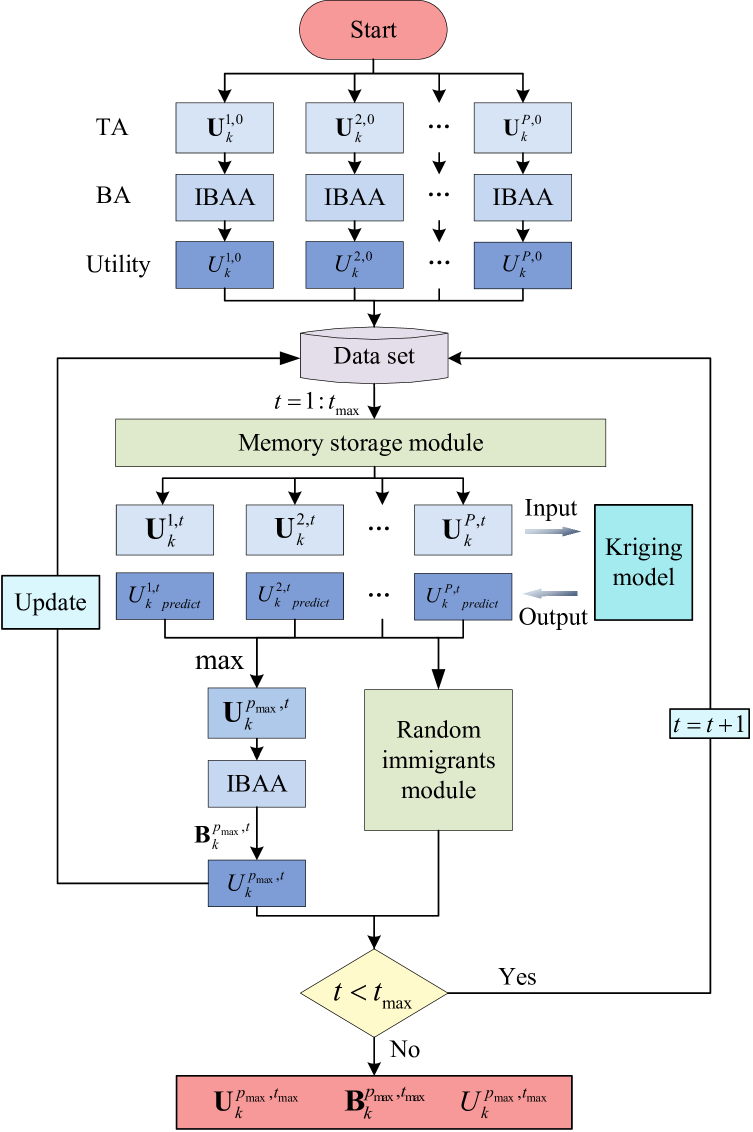}}
	\caption{Flow chart of two-step dynamic hybrid algorithm based on Kriging model at the $k$th frame (TA: task assignment schemes, BA: bandwidth allocation schemes).}
	\label{4n}
\end{figure}
	
%	In the next section, a two-step dynamic hybrid algorithm based on Kriging model is proposed to solve this combinatorial optimization problem.
	
	\begin{remark}
	From Eq. \eqref{19} Eq. \eqref{1} and Eq. \eqref{12}, we find that to evaluate the reward function $R(\mathcal{C}_{m,k})$, $B_{m}$ and ${\rm JSR}_{n,m,k}$ need to be acquired. \textcolor{black}{
		However, it is difficult for UAVs to acquire $B_{m}$ and ${\rm JSR}_{n,m,k}$ because of the adversarial relationship between UAVs and FAR network. The ${\rm JSR}_{n,m,k}$ of UAVs could be estimated\cite{wyh} and $B_{m}$ could be acquired by reconnaissance. 
	There are some errors when estimating $B_{m}$.} From Eq. \eqref{19} and Eq. \eqref{22}, we can see that the inaccurate estimation of $B_{m}$ will affect the suppressive jamming probability $p_{n,m,k}$ and the bandwidth allocation of UAVs $B_{n,m,k}$, thus affecting the suppressive jamming utility. 
	We test the impact of different estimation errors of $B_{m}$ on utility in Section V.
	\end{remark}

	%
	%Therefore, the proposed algorithm will focus on three aspects in the following section.
	%
	%
	%\begin{itemize}
	%	\item The proposed algorithm can find feasible task assignment and bandwidth allocation with maximum utility;
	%	\item The proposed algorithm can adapt to changing dynamic environment;
	%	\item In the case of limited computing resources, the proposed algorithm can still find feasible task assignment and bandwidth allocation with maximum utility.
	%\end{itemize}

	\section{TWO-STEP DYNAMIC HYBRID ALGORITHM BASED ON KRIGING MODEL}
	
	\subsection{Overall Framework}

	As mentioned above, the problem of joint task assignment and bandwidth allocation of UAVs could be formulated as a D-MIP problem with limited resources. To solve this combinatorial optimization problem, a two-step dynamic hybrid algorithm based on Kriging model is proposed in this section. The algorithm has two steps respectively, which are the outer task assignment algorithm (OTAA, i.e., Algorithm \ref{alg:alg11}) and the inner bandwidth allocation algorithm (IBAA, i.e., Algorithm \ref{alg:3}). The flow chart of the overall algorithm is shown in Fig. \ref{4n}. 
	Next, the specific steps of OTAA and IBAA are introduced in detail separately.
	%\begin{figure*}[t]
	%	\centerline{\includegraphics[width=25pc]{liucheng.pdf}}
	%	\caption{Flow chart of two-step dynamic hybrid algorithm.}
	%	\label{4n}
	%\end{figure*}
	%\begin{figure}[t]
	%	\centerline{\includegraphics[width=20pc]{liucheng.pdf}}
	%	\caption{Flow chart of two-step dynamic hybrid algorithm based on Kriging model.}
	%	\label{4n}
	%\end{figure}
	%\begin{figure*}[t]
	%	\centering
	%	\includegraphics[scale=0.16]{liu1.pdf}
	%	\caption{Flow chart of two-step dynamic hybrid algorithm.}
	%	\label{4}
	%\end{figure*}

	%\begin{figure*}[!htbp]
	%	\centerline{\includegraphics[width=10pc]{liu.pdf}}
	%	\caption{The flow chartof two-step dynamic hybrid algorithm.}
	%	\label{fig3}
	%\end{figure*}
	
	\subsection{Outer Task Assignment Algorithm}
	The OTAA uses the framework of GA to efficiently handle discrete task assignment variables, and integrates a memory storage module and a random immigrants module to slow down the convergence of GA and efficiently use the information from previous environment to adapt to dynamic environment. Additionally, a Kriging model is constructed to predict the utility of different task assignment schemes to save computational resources. The specific implementation steps of OTAA are described below.
	
	Latin hypercube sampling (LHS)\cite{loh1996latin} is applied to generate $P$ individuals as the initial task assignment population before the optimization starts, then the initial task assignment population at the $k$th frame is
	\begin{equation}
		\mathcal{U}_k^0=\left[\mathbf{U}_k^{1,0},\mathbf{U}_k^{2,0},\cdots, \mathbf{U}_k^{p,0},\cdots,\mathbf{U}_k^{P,0}\right]^T,
	\end{equation}
	where $\mathbf{U}_k^{p,0}$ is the initial task assignment matrix defined in Eq. \eqref{111111} of the $p$th individual in $\mathcal{U}_k^0$ at the $k$th frame, which is given by
	\begin{equation}
		\mathbf{U}_k^{p,0}=\left[\mathbf{u}_{1,k}^{p,0},\mathbf{u}_{2,k}^{p,0},\cdots,\mathbf{u}_{N,k}^{p,0}\right]^T,
	\end{equation}
	where $\mathbf{u}_{n,k}^{p,0}$ stands for the initial task assignment vector of UAV $n$ in $\mathbf{U}_k^{p,0}$, which can be obtained according to Eq. \eqref{7}, Eq. \eqref{9} and Eq. \eqref{10}.

	After $\mathbf{U}_k^{p,0}$ is determined, the problem of joint task assignment and bandwidth allocation in Eq. \eqref{27} becomes a constrained optimization problem, which uses Algorithm \ref{alg:3} to find the optimal bandwidth allocation scheme $\mathbf{B}_k^{p,0}$ and then obtains the utility, denoted as $U_k^{p,0}$. Then, the utility set of initial task assignment population $\mathcal{U}_k^0$ is
	\begin{equation}
		U(\mathcal{U}_k^0)=\left[U_k^{1,0},U_k^{2,0},\cdots, U_k^{p,0},\cdots,U_k^{P,0})\right]^T.
	\end{equation}

	Then, $\mathcal{U}_k^0$ and $U(\mathcal{U}_k^0)$ are stored in the data set, and Kriging model is established with the data set to represent the relationship between task assignment and utility, that is
	% \begin{equation}
		% 	\begin{bmatrix} U(\mathbf{U}_k^{1,0})\\ 
			% 		U(\mathbf{U}_k^{2,0})\\
			% 		\vdots\\
			% 		U(\mathbf{U}_k^{P,0})\end{bmatrix}
		%=K\begin{bmatrix} \mathbf{U}_k^{1,0}\\ 
			%	\mathbf{U}_k^{2,0}\\
			%	\vdots\\
			%	\mathbf{U}_k^{P,0}\end{bmatrix}
		%\end{equation}
		\begin{equation}
			y=K(\mathbf{U}_k^{p,0}),
		\end{equation}
		where $K(\cdot)$ is the established Kriging model. For each $\mathbf{U}_k^{p,0} (p=1,2,\cdots,P)$ plugged into the Kriging model, $y$ is equal to the corresponding utility $U(\mathbf{U}_k^{p,0})$. 
		The Kriging model can predict the utility of different task assignment schemes without searching for feasible bandwidth allocation schemes, which greatly saves computational resources.
		After the Kriging model is established, the iterative process is carried out, and the memory storage module is entered first.
		\subsubsection{Memory storage module} 
		To store useful information for the current environment, $r_1\cdot P$ individuals are randomly generated as the initial memory task assignment population, and $r_1$ is generally set to 0.1. The initial memory task assignment population at the $k$th frame is
		\begin{equation}
			\mathcal{M}_k^0=\left[\mathbf{U}_k^{1,0},\mathbf{U}_k^{2,0},\cdots,\mathbf{U}_k^{r_1\cdot P,0}\right]^T,
		\end{equation}
		the utility set of $\mathcal{M}_k^0$ is obtained by using Algorithm \ref{alg:3}. The memory update time is set to $t_m$. During the $t$th iteration, $t=1,2,\cdots,t_{max}$, the memory is updated when $t=t_m$. Then, $t_m$ is reset
		\begin{equation}
			\label{337}
			t_m=t+rand(a,b),
		\end{equation}
		where $rand(a,b)$ represents the random generation of an integer from $a$ to $b$. Typically, $a$ could be generally set to 5 and $b$ could be generally set to 10. The memory population $\mathcal{M}_k^t$ is updated according to most similar memory updating strategy \cite{branke1999memory}. 
		\textcolor{black}{If there are randomly initialised individuals in $\mathcal{M}_k^t$, the task assignment with maximum utility in the data set replaces one of them at random. If there are no randomly initialised individuals in $\mathcal{M}_k^t$, it replaces the closest individual in $\mathcal{M}_k^t$ if it is better.}
		
		Next, the task assignment population $\mathcal{U}_k^t$ is updated by crossover and mutation operation of GA\cite{Holland1975AdaptationIN}.
		%A random number $rand_c$ is generated on the interval [0,1], and the crossover probability $p_c$ is set. If $rand_c<p_c$, then two adjacent individuals are crossed by the following formula:
		%\begin{equation}
		%	\label{33}
		%	\mathbf{U}_k^{p,t}=\left[\mathbf{u}_{1,k}^{p,t},\cdots,\mathbf{u}_{n_{rand_c},k}^{p+1,t}\cdots,\mathbf{u}_{N,k}^{p+1,t}\right]^T,
		%\end{equation}
		%\begin{equation}
		%	\label{34}
		%	\mathbf{U}_k^{p+1,t}=\left[\mathbf{u}_{1,k}^{p+1,t},\cdots,\mathbf{u}_{n_{rand_c},k}^{p,t}\cdots,\mathbf{u}_{N,k}^{p,t}\right]^T,
		%\end{equation}
		%where $n_{rand_c}$ is an integer randomly chosen from $\left[1,N\right]$. If $rand_c>p_c$, then two adjacent individuals do not cross.
		%A random number $rand_m$ is generated on the interval $[0,1]$, and the mutation probability $p_m$ is set. If $rand_m<p_m$, an integer $n_{rand_m}$ is chosen from $\left[1,N\right]$ randomly, mutation is carried out on $\mathbf{u}_{n_{rand_m},k}^{p,t}$ and a new task assignment of UAV $n_{rand_m}$ is generated according to \eqref{9} and \eqref{10}. The new task assignment $\mathbf{{u}'}_{n_{rand_m},k}^{p,t}\neq\mathbf{{u}}_{n_{rand_m},k}^{p,t}$.
		%\begin{equation}
		%	\label{35}
		%	\mathbf{U}_k^{p,t}=\left[\mathbf{u}_{1,k}^{p,t},\cdots,\mathbf{u}_{n_{rand_m},k}^{p,t}\cdots,\mathbf{u}_{N,k}^{p,t}\right]^T,\mathbf{u}_{n_{rand_m},k}^{p,t}\neq\mathbf{u}_{n_,k}^{p,t}
		%\end{equation}
		%where $n_{rand_m}$ is an integer randomly chosen from $\left[1,N\right]$. 
		After the new task assignment population is updated, each individual $\mathbf{U}_k^{p,t}$ in $\mathcal{U}_k^t$ is input into Kriging model to obtain predicted utility, that is
		\begin{equation}
			\label{36}
			{U_k^{p,t}}_{predict}=K\left(\mathbf{U}_k^{p,t}\right),
		\end{equation}
		where ${U_k^{p,t}}_{predict}$ is the predicted utility of $\mathbf{U}_k^{p,t}$ obtained by Kriging model. The Kriging model can predict utility quickly without evaluating feasible bandwidth allocation, thus greatly saving computing resources. The individual with the maximum predicted utility was taken out and denoted as $\mathbf{U}_{k}^{p_{max},t}$. Then, Algorithm \ref{alg:3} is used to find true utility of $\mathbf{U}_{k}^{p_{max},t}$, denoted as $U_{k}^{p_{max},t}$. $\mathbf{U}_{k}^{p_{max},t}$ and $U_{k}^{p_{max},t}$ are stored in the data set for updating Kriging model. The uniqueness test of data set is required when updating the Kriging model. 
		
				\begin{algorithm}[t]
			\renewcommand{\algorithmicrequire}{\textbf{Input:}}
			\renewcommand{\algorithmicensure}{\textbf{Output:}}
			\caption{Outer Task Assignment Algorithm.}
			\label{alg:alg11}
			\begin{algorithmic}[1] % 控制是否有序号
				\REQUIRE  The initial task assignment population $\mathcal{U}_k^0$=$[\mathbf{U}_k^{1,0},\mathbf{U}_k^{2,0},\cdots,\mathbf{U}_k^{P,0}]^T$, the initial memory population $\mathcal{M}_k^0$=$[\mathbf{U}_k^{1,0},\mathbf{U}_k^{2,0},\cdots,\mathbf{U}_k^{r_1\cdot P,0}]^T$. % input 的内容
				\ENSURE The optimal task assignment $\mathbf{U}_k^{p_{max},t_{max}}$, its bandwidth allocation $B_k^{p_{max},t_{max}}$ and its utility $U_k^{p_{max},t_{max}}$. % output 的内容
				\STATE 	$\textbf{Initialization:}$ Input $\mathcal{U}_k^0$ and $\mathcal{M}_k^0$ to Algorithm \ref{alg:3} to obtain bandwidth allocation and utility $U(\mathcal{U}_k^0)$ and $U(\mathcal{M}_k^0)$;\\
				\FORALL {$k=1,2,\cdots,K$}
				\STATE Store $\mathcal{U}_k^0$ and $U(\mathcal{U}_k^0)$ into data set and establish Kriging model. Set the initial memory update time $t_m$;
				\FORALL{$t=0,1,\cdots,t_{max}$}
				\IF{$t=t_{m}$}
				\STATE Update $\mathcal{M}_{k}^{t+1}$ according to most similar memory updating strategy. Reset $t_m$ according to \eqref{337};
				\ENDIF
				\STATE Update $\mathcal{U}_k^t$ by crossover and mutation. Set $t=t+1$;
				\STATE Predict utility of each individual in $\mathcal{U}_k^{t+1}$ by Kriging model according to \eqref{36};
				\STATE Replace the worst individuals in $\mathcal{U}_k^{t+1}$;
				\STATE Input the individual with the maximum predicted utility $\mathbf{U}_{k,{max}}^{p,{t+1}}$ into Algorithm \ref{alg:3} to obtain bandwidth allocation and utility $U(\mathbf{U}_{k,{max}}^{p,{t+1}})$;
				\STATE Store $\mathbf{U}_{k,{max}}^{p,{t+1}}$ and $U(\mathbf{U}_{k,{max}}^{p,{t+1}})$ into data set and update Kriging model;
				\ENDFOR
				\STATE Set $k=k+1$;
				\STATE Initialize $P$ individuals by LHS and merge with $\mathcal{M}_{k}^{t_{max}}$. Input these individuals to Algorithm \ref{alg:3} to obtain utility and the top $P$ individuals with the highest utility are taken as $\mathcal{U}_{k+1}^0$.
				\ENDFOR
			\end{algorithmic}
		\end{algorithm}
		
		\subsubsection{Random immigrants module} 
		To avoid premature convergence of the outer task assignment algorithm, $r_2$$\cdot$$P$ individuals are generated randomly to replace $r_2$$\cdot$$P$ individuals with the lowest prediction utility in $\mathcal{U}_k^t$. To avoid the replacement individuals disrupting the search progress, especially when the environment is stable, $r_2$ is generally set to 0.1.

		%\subsubsection{Hybrid memory and random immigrants method}
		
		%In dynamic environments, random immigrants GA (RIGA) is a method to maintain population diversity and avoid premature convergence of GA.
		%In general, RIGA randomly replace individuals or replace individuals with minimum utility in the population. The memory enhanced GA (MEGA) stores useful information from the current environment, and the stored information can be reused later in new environments. In this paper, a hybrid method is proposed by integrating the memory and random immigrants methods, and concrete procedure is given in Algorithm \ref{alg:4}.
		\textcolor{black}{The above iterative process will continue until $t=t_{max}$.} The individual with the maximum utility in the data set is the optimal task assignment at the $k$th frame. This individual is denoted as $\mathbf{U}_k^{p_{max},t_{max}}$, its bandwidth allocation scheme is denoted as $\mathbf{B}_k^{p_{max},t_{max}}$, and its utility is denoted as $U_k^{p_{max},t_{max}}$. \textcolor{black}{Thus, the task assignment scheme $\mathbf{U}_k^{p_{max},t_{max}}$, bandwidth allocation scheme $\mathbf{B}_k^{p_{max},t_{max}}$ and utility $U_k^{p_{max},t_{max}}$ of UAVs at the $k$th frame will be output.}

		Before the optimization starts at the $(k+1)$th frame, LHS is applied to generate $P$ individuals. Each individual is input into Algorithm \ref{alg:3} to obtain utility, and the individual in $\mathcal{M}_{k}^{t_{max}}$ is also input into Algorithm \ref {alg:3} to obtain utility at the $(k+1)$th frame. These individuals are merged, and the top $P$ individuals with the highest utility are taken as the new initial task assignment population $\mathcal{U}_{k+1}^0$ at the $(k+1)$th frame. The remaining steps are the same as the previous time.
		At the first frame, since the initial task assignment population is generated through LHS, a larger $t_{max}$ is needed to achieve convergence. However, for $k=2,3,\cdots,K$, the memory population of previous frame could help initialise the task assignment population of the current frame, so $t^k_{max}$ could be set smaller, i.e.,
		\begin{equation}
			\label{11h11}
			t_{max}^{k}=\alpha\cdot t_{max},k=2,3,\cdots,K,
		\end{equation}
		where $0<\alpha<1$ stands for a constant. 
		The concrete procedure of the OTAA is described in Algorithm \ref{alg:alg11}.

		\subsection{Inner Bandwidth Allocation Algorithm}

		The IBAA uses the framework of FROFI\cite{wang2015incorporating} to efficiently handle continuous bandwidth allocation variables with constraints. The specific implementation steps of IBAA are described below.
		
		After determining the task assignment schemes of UAVs, the problem of joint task assignment and bandwidth allocation in Eq. \eqref{27} becomes a constrained optimization problem about bandwidth allocation of UAVs, that is
		\begin{equation}
			\label{37}
			\begin{aligned}
				&\max~U(\mathbf{B}_{k})\\
				&s.t.\left\{
				\begin{aligned}
					%\nonumber
					&\sum_{n=1}^{N}B_{n,m,k}=B_m,&&m=1,2,\cdots,M\\
					&B_{n,m,k}\geq0,&&n=1,2,\cdots,N\\
					&B_{n,m,k}\leq B_m,&&n=1,2,\cdots,N
				\end{aligned}
				\right.
			\end{aligned}.
		\end{equation}
		
		This constrained optimization problem has $M$ equality constraints and $2N$ inequality constraints.
		For each $\mathbf{U}_k^{p,t}$, the initial bandwidth allocation population of $Q$ individuals is randomly generated, that is
		\begin{equation}
			\mathcal{B}_k^0=\left[\mathbf{B}_k^{1,0},\mathbf{B}_k^{2,0},\cdots, \mathbf{B}_k^{q,0},\cdots,\mathbf{B}_k^{Q,0}\right]^T,
		\end{equation}
		where $\mathbf{B}_k^{q,0}$ is the initial bandwidth allocation matrix defined in Eq. \eqref{999} of the $q$th individual in $\mathcal{B}_k^0$ at $k$th frame, which is given by
		\begin{equation}
			\mathbf{B}_k^{q,0}=\left[\mathbf{B}_{1,k}^{q,0},\mathbf{B}_{2,k}^{q,0},\cdots,\mathbf{B}_{N,k}^{q,0}\right]^T,
		\end{equation}
		where $\mathbf{B}_{n,k}^{q,0}$ is the initial task assignment vector of UAV $n$, which can be obtained according to Eq. \eqref{8} and Eq. \eqref{11}.

		For each $\mathbf{B}_k^{q,0}$, the initial utility $U(\mathbf{B}_k^{q,0})$ is calculated by Eq. \eqref{1111}. Due to the initial bandwidth allocation is generally an infeasible solution, based on the inspiration of \cite{wang2015incorporating}, the initial degree of constraint violation in this article can be defined as
		\begin{equation}
			\label{42}
			\begin{aligned}
				\begin{aligned}
					G(\mathbf{B}_k^{q,0})&=\sum_{m=1}^{M}\max\left(0,\left|\sum_{n=1}^{N}{B_{n,m,k}^{q,0}-B_m}\right|\right)\\
					&+\sum_{m=1}^{M}\sum_{n=1}^{N}\max\left(0,-B_{n,m,k}^{q,0}\right)\\
					&+\sum_{m=1}^{M}\sum_{n=1}^{N}{\max\left(0,B_{n,m,k}^{q,0}-B_m\right)}
				\end{aligned}
			\end{aligned},
		\end{equation}
		where $G(\mathbf{B}_k^{q,0})$ stands for the initial degree of constraint violation of the $q$th individual, $B_{n,m,k}^{q,0}$ is the initial bandwidth allocation defined in Eq. \eqref{11} of UAV $n$ to FAR $m$ of the $q$th individual, $|\cdot|$ is the absolute value operation. If $\mathbf{B}_k^{q,0}$ satisfies the constraints in Eq. \eqref{37}, then $G(\mathbf{B}_k^{q,0})=0$ and $\mathbf{B}_k^{q,0}$ is a feasible solution. Conversely, $G(\mathbf{B}_k^{q,0})>0$ and $\mathbf{B}_k^{q,0}$ is an infeasible solution.

		During the $g$th ($g=1,2,\cdots,g_{max}$) iteration, the bandwidth allocation population $\mathcal{B}_k^g$ is updated by mutation, crossover and selection operation \cite{storn1997differential}. Two mutation operators are adopted as follows
		\begin{equation}
			\label{43}
			\mathbf{v}_k^{q,g}=\mathbf{B}_k^{q,g}+F\cdot\left(\mathbf{B}_k^{q_1,g}-\mathbf{B}_k^{q,g}\right)+F\cdot\left(\mathbf{B}_k^{q_2,g}-\mathbf{B}_k^{q_3,g}\right),
		\end{equation}
		\begin{equation}
			\label{44}
			\mathbf{v}_k^{q,g}=\mathbf{B}_k^{q_1,g}+F\cdot\left(\mathbf{B}_k^{q_{best},g}-\mathbf{B}_k^{q_1,g}\right)+F\cdot\left(\mathbf{B}_k^{q_2,g}-\mathbf{B}_k^{q_3,g}\right),
		\end{equation}
		where $q_1$, $q_2$, and $q_3$ are mutually different 
		integers chosen from $\left[1,Q\right]$  randomly, $\mathbf{B}_k^{q_{best},g}$ stands for the best 
		bandwidth allocation individual in current population, $F$ stands for the scaling factor, and $\mathbf{v}_k^{q,g}=[\mathbf{v}_{1,k}^{q,g},\mathbf{v}_{2,k}^{q,g},\cdots,\mathbf{v}_{N,k}^{q,g}]$ is the mutant vector.

		\begin{algorithm}[t]
			\renewcommand{\algorithmicrequire}{\textbf{Input:}}
			\renewcommand{\algorithmicensure}{\textbf{Output:}}
			\caption{Inner Bandwidth Allocation Algorithm.}
			\label{alg:3}
			\begin{algorithmic}[1] % 控制是否有序号
				\REQUIRE The task assignment individual $\mathbf{U}_k^{p,t}$. % input 的内容
				\ENSURE The optimal bandwidth allocation $\mathbf{B}_k^{p_{max},t_{max}}$ and its utility ${U}_k^{p_{max},t_{max}}$. % output 的内容
				\STATE 	$\textbf{Initialization:}$ Generate the initial bandwidth allocation population $\mathcal{B}_k^0=[\mathbf{B}_k^{1,0},\mathbf{B}_k^{2,0},\cdots,\mathbf{B}_k^{Q,0}]^T$;\\
				\FORALL {$g=0,1,\cdots,g_{max}$}
				\STATE Generate trial vector $\mathbf{w}_k^{q,g}$ by mutation and crossover operations according to \eqref{43}, \eqref{44} and \eqref{45};
				\STATE Evaluate utility and degree of constraint violation for $\mathbf{w}_k^{q,g}$ according to \eqref{1111}, \eqref{42} ($g=0$) and \eqref{65} ($g=1,\cdots,g_{max}$);
				\STATE Compare $\mathbf{B}_k^{q,g}$ with $\mathbf{w}_k^{q,g}$ by the feasibility rule according to \eqref{46} and \eqref{47}. Set $g=g+1$;
				\IF{$G(\mathbf{w}_k^{q,g})>G(\mathbf{B}_k^{q,g})$ {\rm and $U(\mathbf{w}_k^{q,g})>U(\mathbf{B}_k^{q,g}$)}}
				\STATE Put $\mathbf{w}_k^{q,g}$ into replacement population $A$;
				\ENDIF
				\STATE Replace some individuals in $\mathcal{B}_k^{g+1}$ with the individuals in $A$ according to the replacement mechanism;
				\IF{all the individuals in $\mathcal{B}_k^{g+1}$ are infeasible solutions}
				\STATE Implement the mutation strategy.
				\ENDIF
				\ENDFOR
			\end{algorithmic}
		\end{algorithm}	
		
		The $\mathbf{B}_k^{q,g}$ and mutant vector $\mathbf{v}_k^{q,g}$ generate the trial 
		vector $\mathbf{w}_k^{q,g}=[\mathbf{w}_{1,k}^{q,g},\mathbf{w}_{2,k}^{q,g},\cdots,\mathbf{w}_{N,k}^{q,g}]$ by crossover operation. The binomial crossover is implemented below
		\begin{equation}
			\label{45}
			\mathbf{w}_{n,k}^{q,g}=\left\{
			\begin{aligned}
				%\nonumber
				&\mathbf{v}_{n,k}^{q,g}, 
				&&{\rm if}~rand_{n}\leq CR~{\rm or}~n=n_{rand}\\
				&\mathbf{B}_{n,k}^{q,g}, &&{\rm otherwise} \\
			\end{aligned},
			\right.
		\end{equation}
		where $n=1,2,\cdots,N$, $rand_n$ stands for a uniformly distributed random number on the interval $[0,1]$ and regenerated for each $n$, $n_{rand}$ stands for an integer chosen from $\left[1,N\right]$ randomly, and $CR$ stands for the crossover control parameter. Then, the utility of $\mathbf{B}_k^{q,g}$ and $\mathbf{w}_k^{q,g}$ are calculated by Eq. \eqref{1111} and the degree of constraint violation could be obtained by 
		\begin{equation}
			\begin{aligned}
				\label{65}
				\begin{aligned}
					G(\mathbf{B}_k^{q,g})&=\sum_{m=1}^{M}\max\left(0,\left|\sum_{n=1}^{N}{B_{n,m,k}^{q,g}-B_m}\right|\right)\\
					&+\sum_{m=1}^{M}\sum_{n=1}^{N}\max\left(0,-B_{n,m,k}^{q,g}\right)\\
					&+\sum_{m=1}^{M}\sum_{n=1}^{N}{\max\left(0,B_{n,m,k}^{q,g}-B_m\right)}
				\end{aligned}
			\end{aligned},
		\end{equation}
		where $G(\mathbf{B}_k^{q,g})$ stands for the degree of constraint violation of the $q$th individual in the $g$th iteration, $B_{n,m,k}^{q,g}$ is the bandwidth allocation defined in Eq. \eqref{11} of UAV $n$ to FAR $m$ of the $q$th individual in the $g$th iteration.

			%	In this table, the distance is measured in km, the velocity is measured in m/s, the angular velocity is measured in rad/s, the acceleration is measured in m/s$^2$.

		The $\mathbf{B}_k^{q,g}$ is compared with its trial vector $\mathbf{w}_k^{q,g}$ and the better one will be selected for the next generation according to the following formula, that is 
		\begin{equation}
			\label{46}
			\mathbf{B}_k^{q,g+1}=\left\{
			\begin{aligned}
				%\nonumber
				&\mathbf{w}_k^{q,g}, 
				&&{\rm if}~G(\mathbf{w}_k^{q,g})< G(\mathbf{B}_k^{q,g})\\
				&\mathbf{B}_k^{q,g},&&{\rm if}~G(\mathbf{w}_k^{q,g})> G(\mathbf{B}_k^{q,g})\\
			\end{aligned}.
			\right.
		\end{equation}
		
		It can be seen that the individual with small degree of constraint violation is selected for the next generation. When $G(\mathbf{w}_k^{q,g})=G(\mathbf{B}_k^{q,g})$, the utility of the two individuals is compared, that is
		\begin{equation}
			\label{47}
			\mathbf{B}_k^{q,g+1}=\left\{
			\begin{aligned}
				%\nonumber
				&\mathbf{w}_k^{q,g}, 
				&&{\rm if}~U(\mathbf{w}_k^{q,g})\geq U(\mathbf{B}_k^{q,g})\\
				&\mathbf{B}_k^{q,g}, &&{\rm if}~U(\mathbf{w}_k^{q,g})< U(\mathbf{B}_k^{q,g})\\
			\end{aligned}.
			\right.
		\end{equation}
		
        					\begin{table}[t]
        	\centering
        	\caption{Simulation parameter settings}
        	\label{tq1}
        	\begin{tabular}{c|c}
        		\toprule 
        		\hline
        		Location of FARs (km)& (0,15) (15,0) (0,0) \\
        		\hline
        		Transmitting power of FARs & $P_t^m=10$kW  \\
        		\hline
        		Working bandwidth of FARs & $B_m=500$MHz  \\
        		\hline
        		Main lobe gain of FARs & $G_r^m=40$dB  \\	
        		\hline
        		Main lobe width of FARs & $\theta_{0.5}=1$rad  \\
        		\hline
        		Interval between adjacent frames & $\Delta T=5$s  \\
        		\hline
        		\multirow{3}*{Location of UAVs (km)}&(24,1) (22,3) (20,5) (18,7)\\&(16,9) (14,11) (12,13) (10,15)\\&(8,17) (6,19) (4,21) (2,23)\\
        		\hline
        		Transmitting power of UAVs & $P_j^n=1$W  \\
        		\hline
        		Main lobe gain of UAVs  & $G_j^n=5$dB  \\
        		\hline
        		Polarization loss of suppressive& \multirow{2}{*}{$\gamma_j=0.5$}  \\
        		jamming signal to FAR antenna&\\
        		\hline
        		Gain constant of FAR&\multirow{2}{*}{$\xi=0.313$}\\
        		to suppressive jamming signal &   \\
        		\hline
        		RCS of high-value target & $\sigma=25$m$^2$  \\
        		\hline
        		\multirow{2}*{Initial state of high-value target}&($x_0$,$v_x$,$y_0$,$v_y$)\\&=(69km,-240m/s,59km,-240m/s)\\
        		%		\hline
        		%		\multirow{2}*{Initial state of CA model}&($x_0$,$v_x$,$a_x$,$y_0$,$v_y$,$a_y$,)\\&=(75,-240,-50,65,-240,-50)  \\
        		%		\hline
        		%		\multirow{2}*{Initial state of CT model}&($x_0$,$v_x$,$y_0$,$v_y$,$\omega_a$)\\
        		%		&=(75,-240,65,240,1)\\
        		\hline
        		\multirow{2}*{Process noise}&$x$ direction: Gaussian of $N(0,3^2)$\\
        		&$y$ direction: Gaussian of $N(0,3^2)$\\
        		\hline
        		Requirement of JSR & ${\rm JSR}^o=1e^{-5}$  \\
        		\hline
        		Tolerance factor of suppressive&\multirow{2}{*}{$L=1$}  \\
        		jamming effect function $f(\cdot)$&\\
        		\hline
        		%		Cost factor & $\lambda=0.0001$  \\	
        		%		\hline
        		\bottomrule
        	\end{tabular}
        \end{table}
		
		The above proposed selection strategy satisfies the feasibility rule\cite{deb2000efficient}. However, the feasibility rule prefers constraints to objective function, and the information of objective function is not well used. To utilise the information of the constraints and objective function, FROFI algorithm is proposed \cite{wang2015incorporating}. Unlike feasibility rule, FROFI has a replacement mechanism and a mutation strategy. When $G(\mathbf{w}_k^{q,g})>G(\mathbf{B}_k^{q,g})$ and $U(\mathbf{w}_k^{q,g})>U(\mathbf{B}_k^{q,g})$, $\mathbf{w}_k^{q,g}$ is deposited in the replacement population $A$. 
%		It is worth noting that $\mathbf{w}_k^{q,g}$ is an infeasible solution with high utility. 
		After the new generation $\mathcal{B}_k^{g+1}$ is generated through the above mutation, crossover and selection operations, the individual in $\mathcal{B}_k^{g+1}$ is replaced by the individual in $A$ through a certain replacement mechanism. When all the solutions in the population are infeasible, a mutation strategy is implemented. The replacement mechanism can make good use of the infeasible solutions with high utility, and the mutation strategy can jump out of the infeasible region. The validity of replacement mechanism and mutation strategy are verified in \cite{wang2015incorporating}.

		\textcolor{black}{The above iterative process will continue until $g=g_{max}$.} The final bandwidth allocation scheme of $\mathbf{U}_k^{p,t}$ at the $k$th frame will be outputted, denoted as $\mathbf{B}_k^{p_{max},t_{max}}$. Its utility is denoted as ${U}_k^{p_{max},t_{max}}$. The concrete procedure of the IBAA is given in Algorithm \ref{alg:3}.

			\section{SIMULATION RESULTS}

			\textcolor{black}{To validate the effectiveness of the above proposed algorithm, we analyse it from various aspects such as convergence and jamming performance.
		   The specific parameters are shown in Table \ref{tq1}.}

			\subsection{Convergence Analysis}
			
						\begin{table}[t]
				\centering
				\caption{Utility and standard deviation under different initial population sizes of Kriging model}
				\label{t1}
				%\resizebox{\linewidth}{!}{
					\begin{tabular}{c|c|c}% 通过添加 | 		
						\toprule
						\hline 
						Initial population size&Utility&Standard deviation\\ 
						\hline
						$P=2$&$U=0.8421$&$\pm$0.1650\\
						\hline
						$P=6$&$U=0.9478$&$\pm$0.0021\\
						\hline
						$P=10$&$U=0.9463$&$\pm$0.0022\\
						\hline
						$P=20$&$U=0.9448$&$\pm$0.0182\\
						\hline
						\bottomrule
					\end{tabular}
					%}
			\end{table}	
			
			\textcolor{black}{We first analyse the convergence of the proposed algorithm at the first frame. The weights are set to $\omega_1=\omega_2=\omega_3=\frac{1}{3}$, the cost factor could be set to $\lambda=0$.} We discuss the influence of the initial population size on Kriging model at the first frame. The initial population size of Kriging model is taken as 2, 6, 10 and 20 respectively, the number of iterations of outer task assignment algorithm $t^1_{max}=t_{max}+1=500$ and the total computing resources $\mathcal{O}_1=1\times10^8$. \textcolor{black}{The utility and standard deviation are obtained from the average of more than 100 independent tests,} as shown in Table \ref{t1}. 
			We find that when the initial population size $P=2$, the utility is the smallest and the standard deviation is largest, indicating that the initial population size is too small to affect the accuracy of Kriging model. 
			However, the utility does not keep increasing as $P$ increases. This is because computational resources are also consumed when the initial population is too large, leading to insufficient computational resources for subsequent iterations, resulting in small utility.
			In the following simulation, the initial population size of the following simulation is $P=6$.

\begin{figure}[t]
				\centerline{\includegraphics[width=19pc]{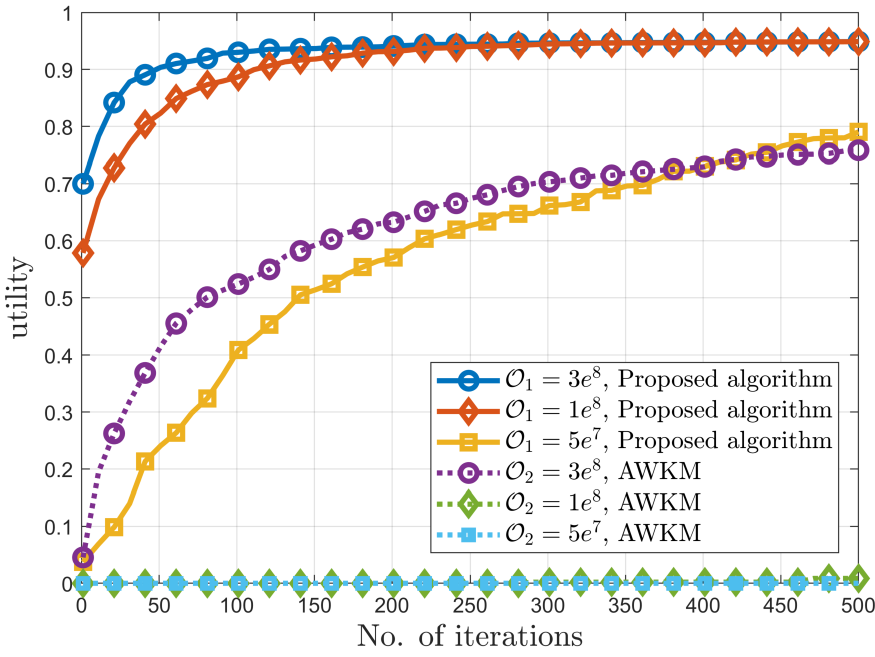}}
				\caption{The curves of utility of proposed algorithm and AWKM under different total computing resources at the first frame. }
				\label{5}
			\end{figure}

			To verify the effectiveness of Kriging model in saving computing resources, the proposed algorithm is compared with algorithm without Kriging model (AWKM). The total computing resources of AWKM are denoted as $\mathcal{O}_2$.
%			To be fair, the total computing resources of the two algorithms should be the same, that is, $\mathcal{O}_1=\mathcal{O}_2$. For a more comprehensive comparison, we also test the above two algorithms using different total computing resources. 
			The curves of total utility of proposed algorithm and AWKM under different total computing resources at the first frame are shown in Fig. \ref{5}. \textcolor{black}{The curves are obtained from the average of more than 100 independent tests.}
			We find that when the total computing resource $\mathcal{O}_1=1e^8$, the curve of the proposed algorithm converges at 400 iterations. 
			When $\mathcal{O}_1=3e^8$, the utility of the proposed algorithm hardly does not increase significantly. 
			This is because when  $\mathcal{O}_1=1e^8$, the computational resources are already sufficient and continuing to add computational resources will result in redundancy. When the total computing resource $\mathcal{O}_1=5e^7$, the curve of the proposed algorithm does not converge and the utility decreases significantly, indicating that the total computing resources are insufficient. Therefore, we set $t^1_{max}=400$ and $\mathcal{O}_1=1e^8$ for the proposed algorithm at first frame in the next simulations.
			
			Under the same total computing resources, the utility of the proposed algorithm is much higher than that of AWKM. When $\mathcal{O}_2=5e^7$ and $\mathcal{O}_2=1e^8$, AWKM cannot even find a feasible task assignment and bandwidth allocation scheme. When $\mathcal{O}_2=3e^8$, the utility of AWKM increases significantly, but it has not reached convergence, indicating that the computing resources are insufficient. The utility of AWKM is highly dependent on the total computing resources. However, the proposed algorithm can find task assignment and bandwidth allocation schemes with higher utility in less computing resources.

\begin{figure}[t]
				\centerline{\includegraphics[width=19pc]{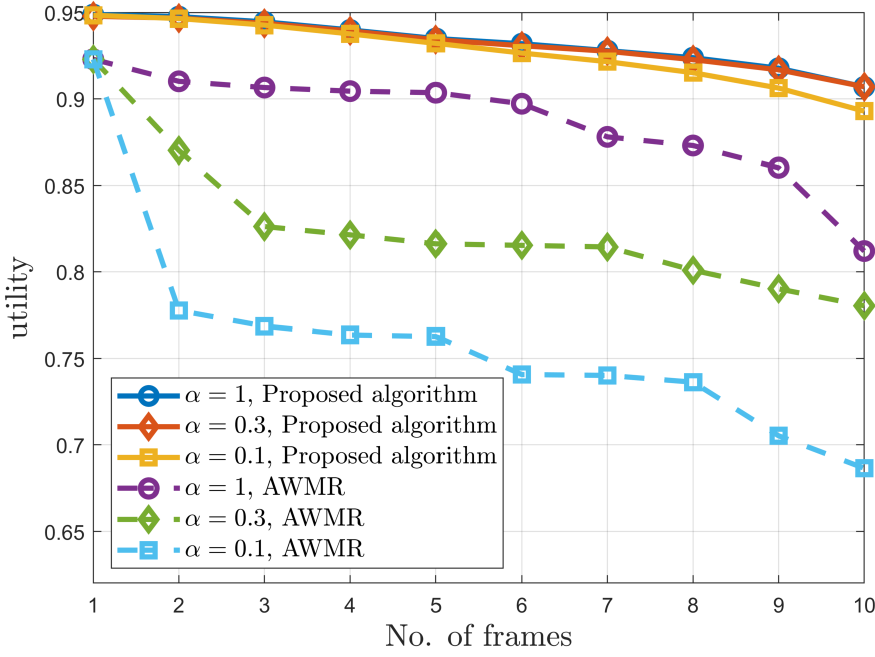}}
				\caption{The curves of utility of proposed algorithm and AWMR under different $\alpha$.}
				\label{dt}
			\end{figure}

			\textcolor{black}{Then, we analyse the convergence of the proposed algorithm in the next frames (i.e., $k=2,3,\cdots,K$).} To verify the effectiveness of memory storage module and random immigrants module, we compare the proposed algorithm with the algorithm without memory storage module and random immigrants module (AWMR). To be fair, the weight coefficients are $\omega_1=\omega_2=\omega_3=\frac{1}{3}$ and the cost factor is $\lambda=0$ of the two algorithms. 
			%			For a more comprehensive comparison, we also test the above two algorithms using different $\alpha$ (As shown in Eq. \eqref{11h11}, $\alpha$ decides the maximum number of iterations at the $k$th frame, $k=2,3,\cdots,K$). 
			The curves of utility of proposed algorithm and AWMR at different frames under different $\alpha$ are shown in Fig. \ref{dt}. \textcolor{black}{The curves are obtained from the average of more than 100 independent tests.} We observe that the utility of the proposed algorithm is roughly equal when $\alpha=0.3$ and when $\alpha=1$, indicating that the proposed algorithm reaches convergence when $\alpha=0.3$. When $\alpha=0.1$, the utility of the proposed algorithm decreases, indicating that the number of iterations is too small to converge. Thus, the proposed algorithm with $\alpha=0.3$ is set in the next simulations. 
			
			Under the same $\alpha$, the utility of proposed algorithm outperforms that of AWMR. When $\alpha=1$, the utility of the proposed algorithm is slightly larger than that of the AWMR. This is because random immigrants module prevents OTAA from converging prematurely to explore better solutions. When $\alpha=0.3$ and $\alpha=0.1$, the utility of the AWMR decreases significantly from frame 2 to frame 10. This is because the memory storage module can store the optimal solution of the previous frame for the next frame, and help the initialization of next frame. 
			\textcolor{black}{In addition, we find that the utility decreases gradually as the number of frames increases.} This is because as high-value target close to FAR network, the echo signal received by FAR network becomes larger, making the JSR defined in Eq. \eqref{18} smaller and making it more difficult for UAVs to carry out jamming tasks.

			\subsection{Performance Analysis}
			\subsubsection{Comparative experiments}To demonstrate the performance of the proposed algorithm, we compare the proposed algorithm with three task assignment and bandwidth allocation algorithms. The characteristics of the three task assignment and bandwidth allocation algorithms are given as follows.
		\begin{itemize}
			\item \textit{Outer task assignment algorithm-bandwidth equalization (TABE):} The algorithm only considers the optimization of task assignment, that is, the outer task assignment algorithm is used for task assignment of UAVs, while the bandwidth of UAVs that suppress the same FAR is equally divided.
			\item \textit{Task assignment fixation-inner bandwidth allocation algorithm (TFBA):} The algorithm only considers the optimization of bandwidth allocation, that is, the task selection of UAVs are fixed according to the distance from the FAR and the inner bandwidth allocation algorithm is used for bandwidth allocation of UAVs. By calculating the distance between UAVs and FAR network, the first six UAVs suppress FAR 1 and the last six UAVs suppress FAR 2.
			\item \textit{Task assignment fixation-bandwidth allocation equalization (TFBE):} The algorithms do not consider the optimization of task assignment and bandwidth allocation, that is, the task assignment of UAVs is fixed, same as TFBA and the bandwidth of UAVs that suppress the same FAR is equally divided.
		\end{itemize}
		
					\begin{figure*}[t]
			\centering
			\subfloat[]{\includegraphics[width=2.25in]{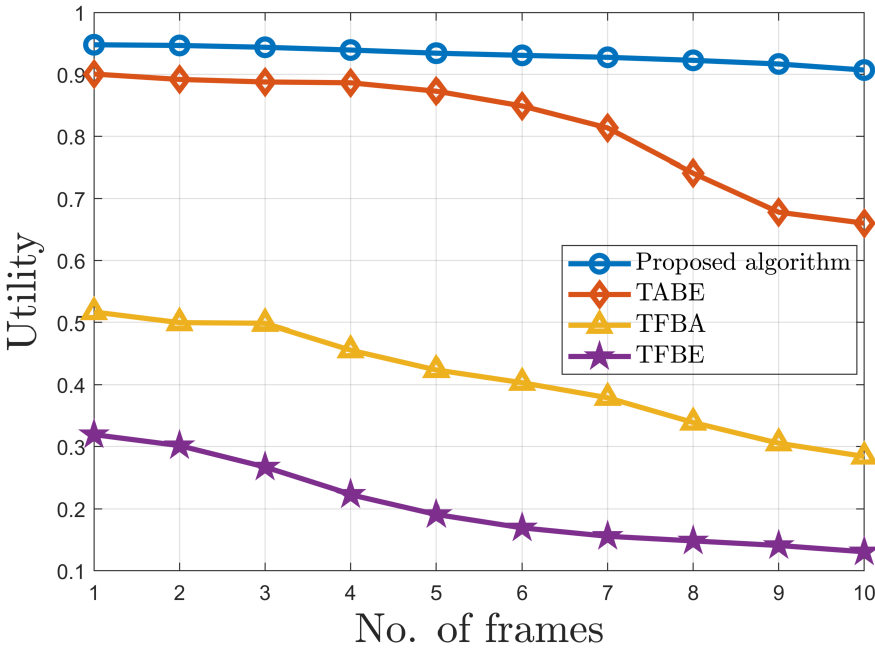}%
				\label{a1}}
			\hfil
			\subfloat[]{\includegraphics[width=2.25in]{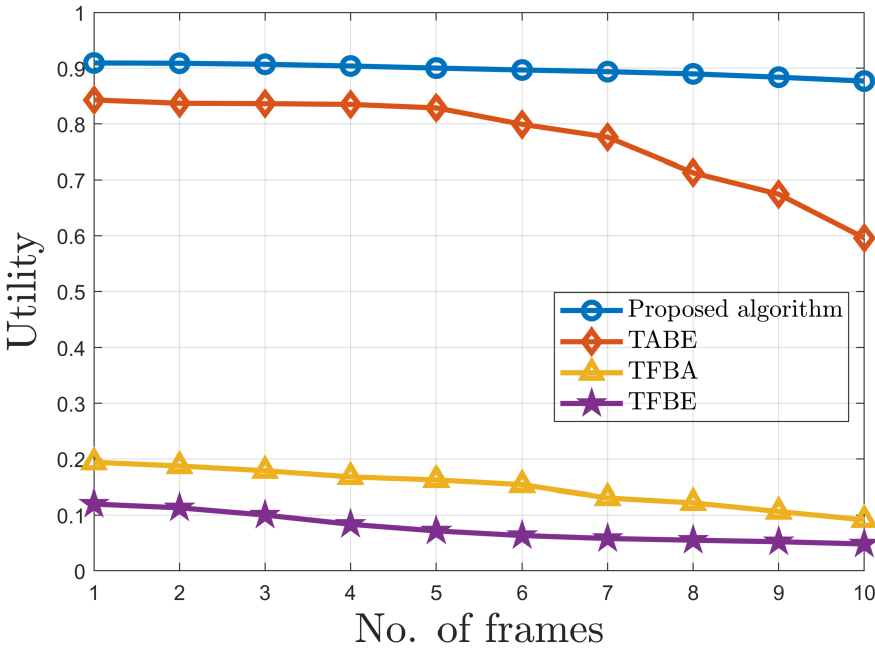}%
				\label{a2}}
			\hfil
			\subfloat[]{\includegraphics[width=2.25in]{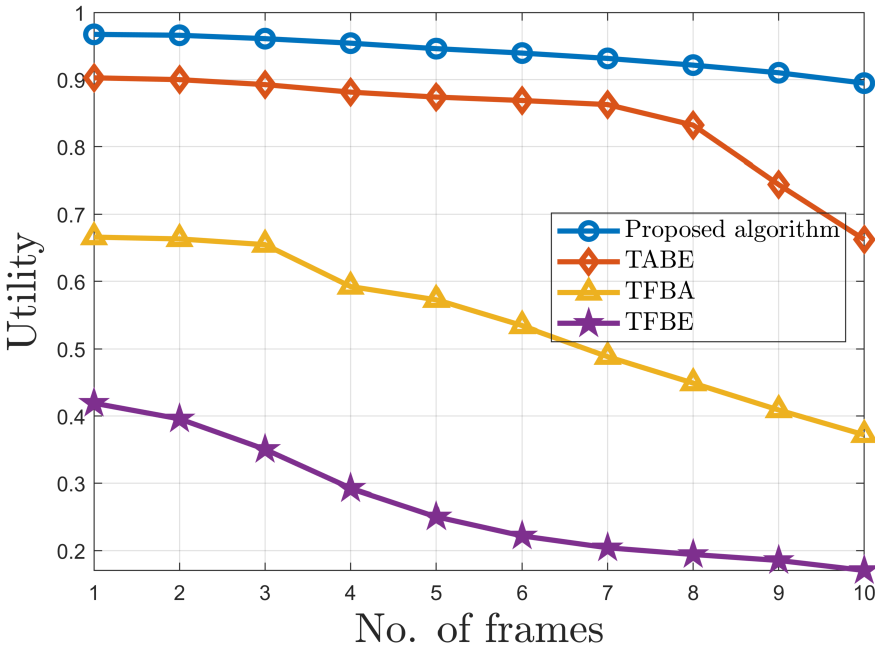}%
				\label{a3}}
			%	\hfil
			%	\subfloat[]{\includegraphics[width=2.2in]{set1 0.1.png}%
				%		\label{a21}}
			%	\hfil
			%	\subfloat[]{\includegraphics[width=2.2in]{set2 0.1.png}%
				%		\label{a22}}
			%	\hfil
			%	\subfloat[]{\includegraphics[width=2.2in]{set3 0.1.png}%
				%		\label{a23}}
			\caption{The curves of utility by different task assignment and bandwidth allocation algorithms. (a) Setting 1; (b) Setting 2; (c) Setting 3.}
			%	; (d) \textbf{Setting 1}, $\lambda=0.1$; (e) \textbf{Setting 2}, $\lambda=0.1$; (f) \textbf{Setting 3}, $\lambda=0.1$.}
		\label{123}
	\end{figure*}
		
			\begin{figure}[t]
			\centerline{\includegraphics[width=19pc]{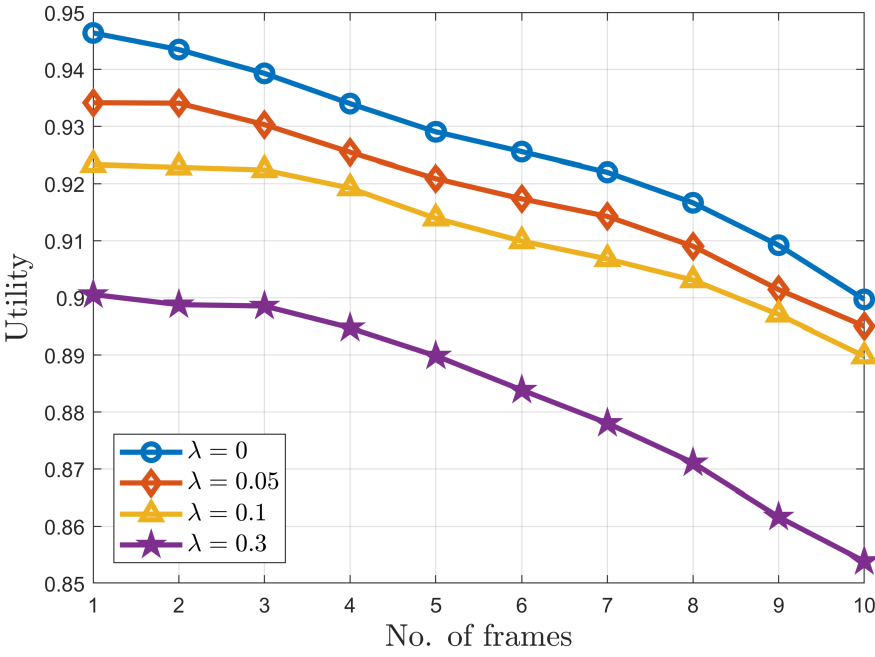}}
			\caption{The curves of utility of proposed algorithm under different $\lambda$.}
			\label{MDP}
		\end{figure}
		
		\textcolor{black}{The weights of the four algorithms are set for three cases, i.e., Setting 1: $\omega_1=\omega_2=\omega_3=\frac{1}{3}$; Setting 2: $\omega_1=\omega_2=\frac{1}{8}, \omega_3=\frac{3}{4}$; Setting 3: $\omega_1=\omega_2=\frac{7}{16}, \omega_3=\frac{1}{8}$). The cost factor is set to $\lambda=0$. 
		The curves of utility by the above four algorithms under different settings are shown in Fig. \ref{123}. The curves are obtained from the average of more than 100 independent tests.}
		We observe that the utility of the proposed algorithm is always maximised. Since both task assignment and bandwidth allocation of UAVs are important to suppressive jamming utility, task fixation or bandwidth equalization will reduce utility. 
		\textcolor{black}{We also observe a decrease in the utility of all four algorithms under Setting 2 compared to the utility under Setting 1 and Setting 3. This is because task 3 is the most difficult of all the tasks, and the UAVs obtain small utility by performing task 3. When the weight $\omega_3$ is set maximum for task 3 (i.e., Setting 2), it increases the difficulty of jamming for UAVs, so the utility of the four algorithms is minimum under Setting 2. In conclusion, the performance of the proposed algorithm is the best compared to the other three task assignment and bandwidth allocation algorithms, proving the effectiveness of the optimisation of task assignment and optimisation of bandwidth allocation. The other three algorithms have less utility due to optimising only task assignment or bandwidth allocation or neither of them.}

		\subsubsection{Effect of cost factor}
		
		%\begin{figure}[t]
		%	\centerline{\includegraphics[width=19pc]{daijia.png}}
		%	\caption{ The curves of utility of proposed algorithm under different $\lambda$.}
		%	\label{daijia}
		%\end{figure}

		\begin{figure*}[t]
			\centering
			\subfloat[]{\includegraphics[width=19pc]{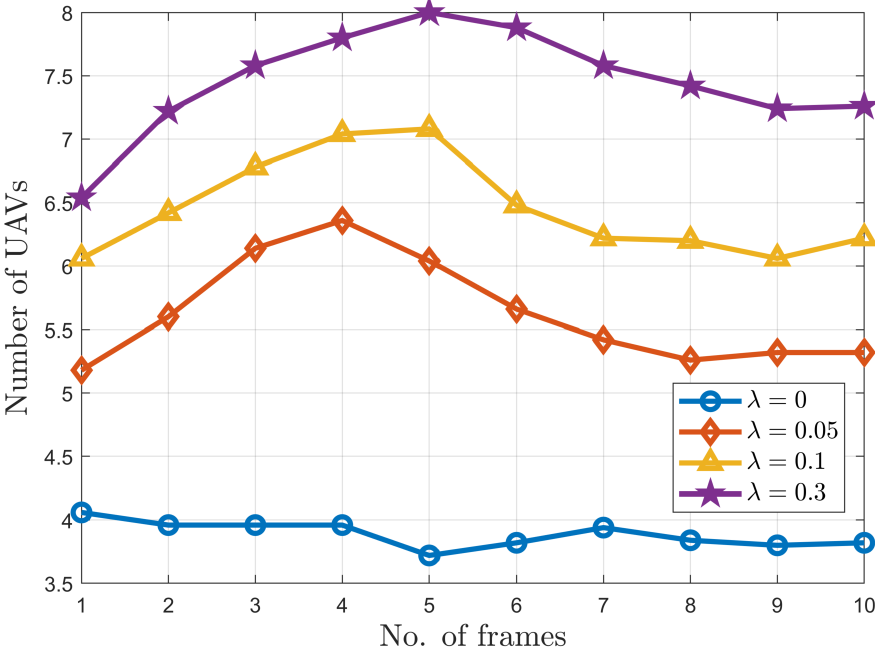}%
				\label{a51}}
			\hfil
			\subfloat[]{\includegraphics[width=19pc]{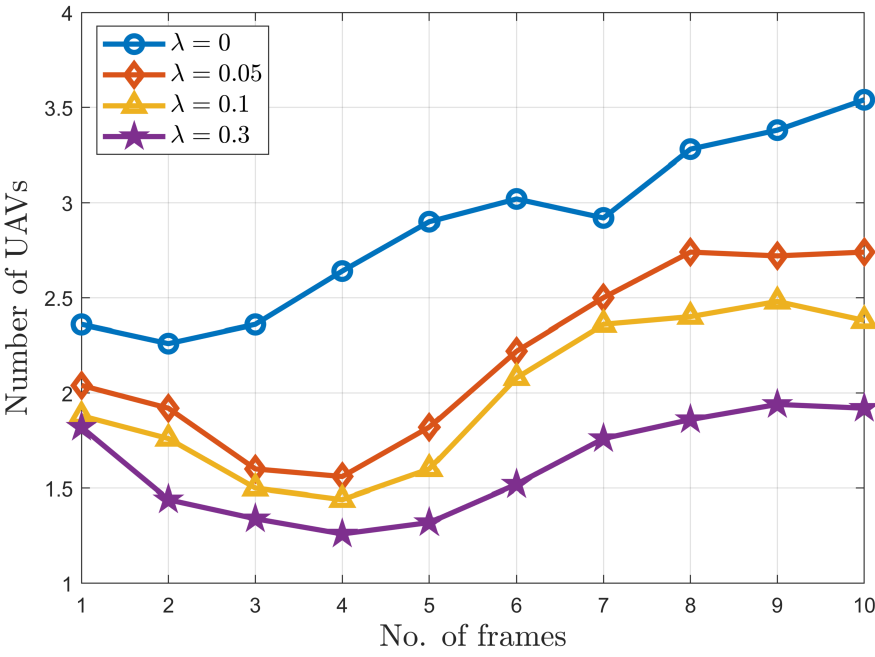}%
				\label{a52}}
			\hfil
			\subfloat[]{\includegraphics[width=19pc]{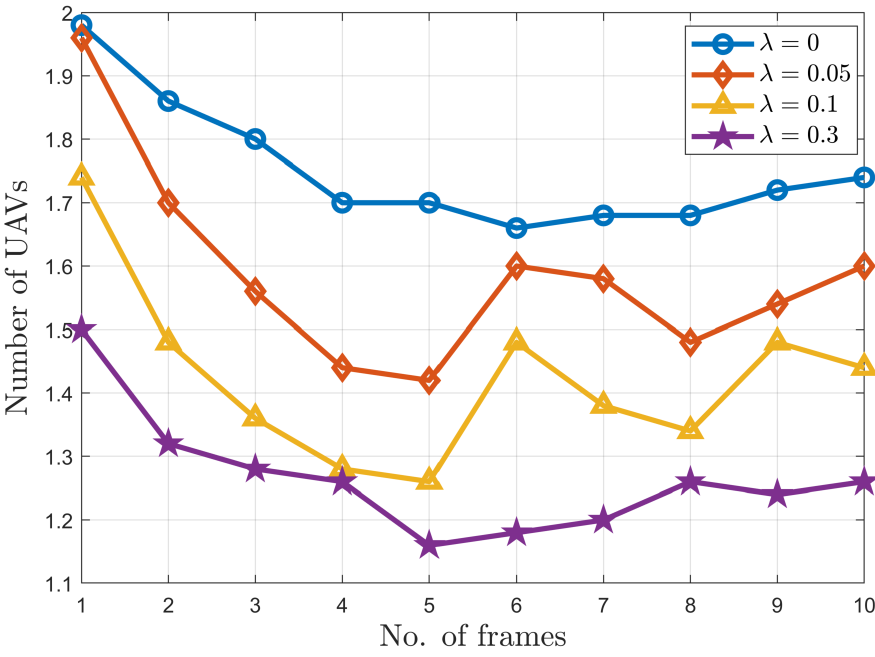}%
				\label{a53}}
			\hfil
			\subfloat[]{\includegraphics[width=19pc]{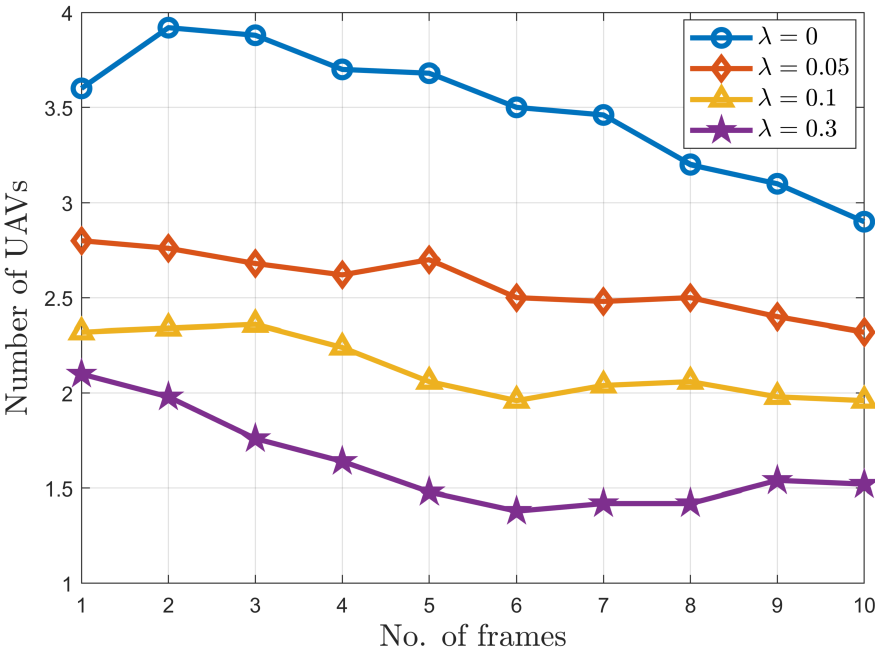}%
				\label{a251}}
			\caption{Number of UAVs for different FARs under different $\lambda$. (a) The number of Non-working UAVs; (b) The number of UAVs assigned to FAR 1; (c) The number of UAVs assigned to FAR 2; (d) The number of UAVs assigned to FAR 3.}
			\label{1234}
		\end{figure*}

		\begin{figure*}[t]
			\centering
			\subfloat[]{\includegraphics[width=2.33in]{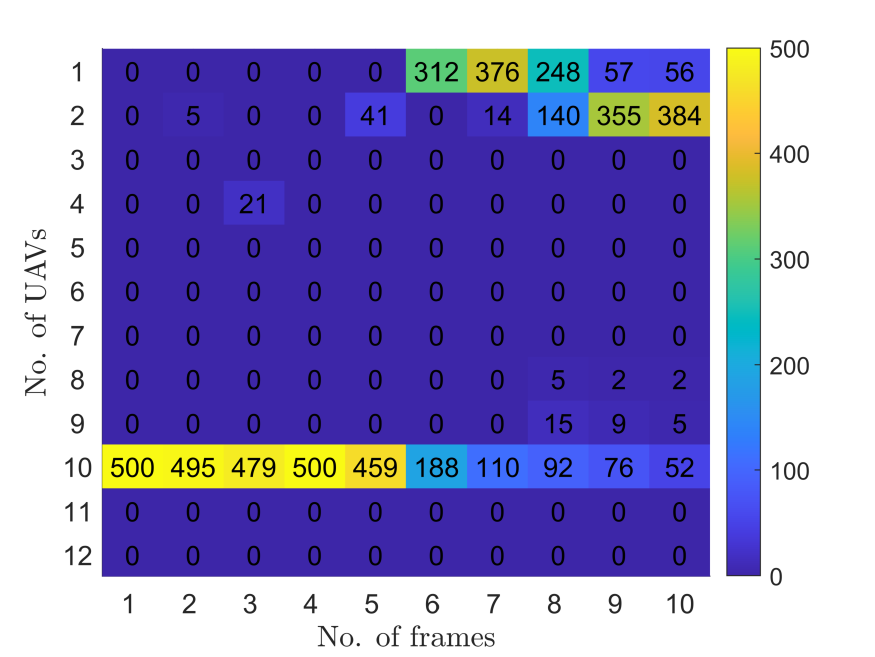}%
				\label{aaa}}
			\hfil
			\subfloat[]{\includegraphics[width=2.33in]{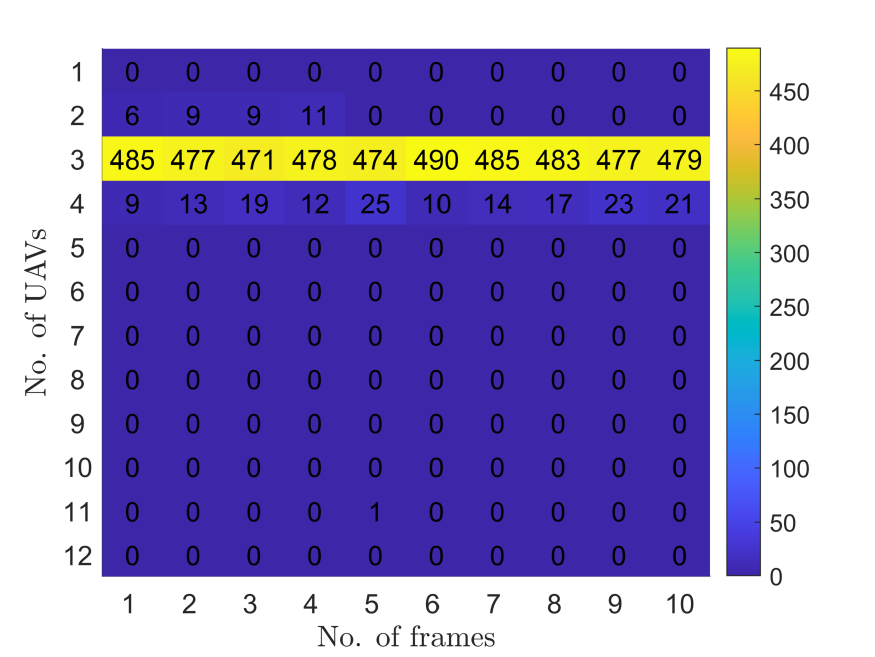}%
				\label{bbb}}
			\hfil
			\subfloat[]{\includegraphics[width=2.33in]{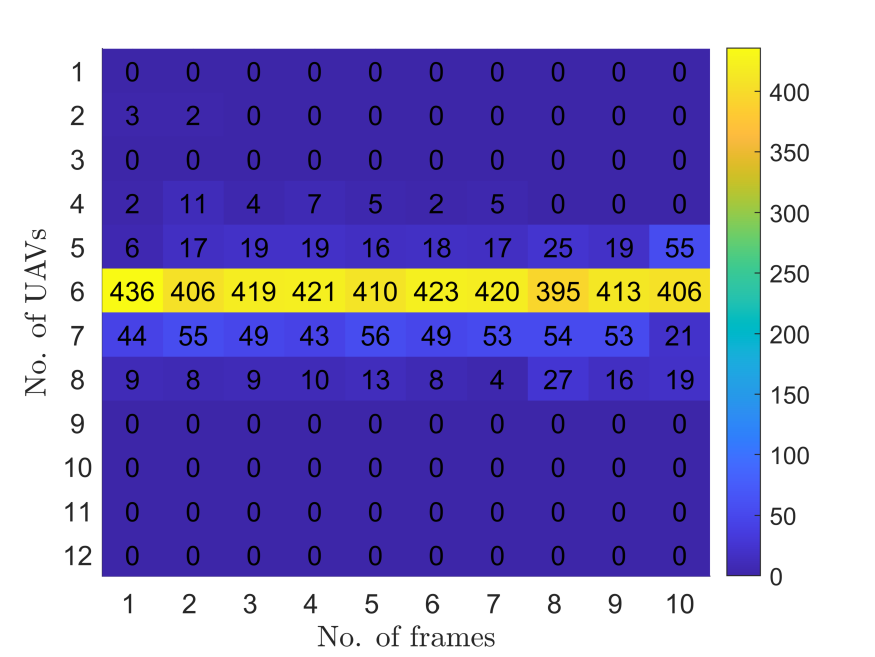}%
				\label{ccc}}
			\caption{The bandwidth allocation of UAVs in different frames when $\lambda=0$. (a) Bandwidth allocation of UAVs to FAR 1; (b) Bandwidth allocation of UAVs to FAR 2; (c) Bandwidth allocation of UAVs to FAR 3.}
			\label{777}
		\end{figure*}
		
		\begin{figure*}[t]
			\centering
			\subfloat[]{\includegraphics[width=2.33in]{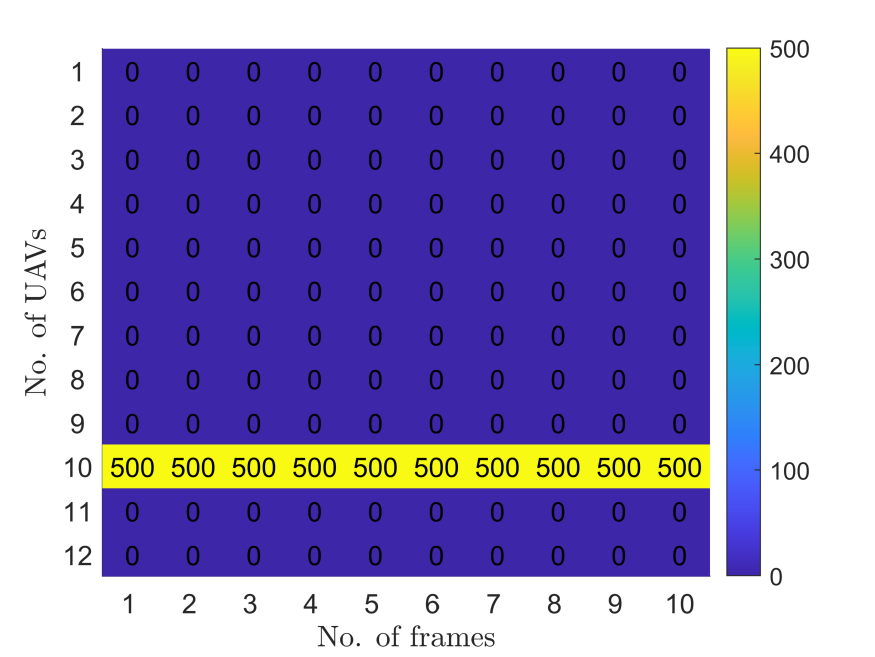}%
				\label{aa}}
			\hfil
			\subfloat[]{\includegraphics[width=2.33in]{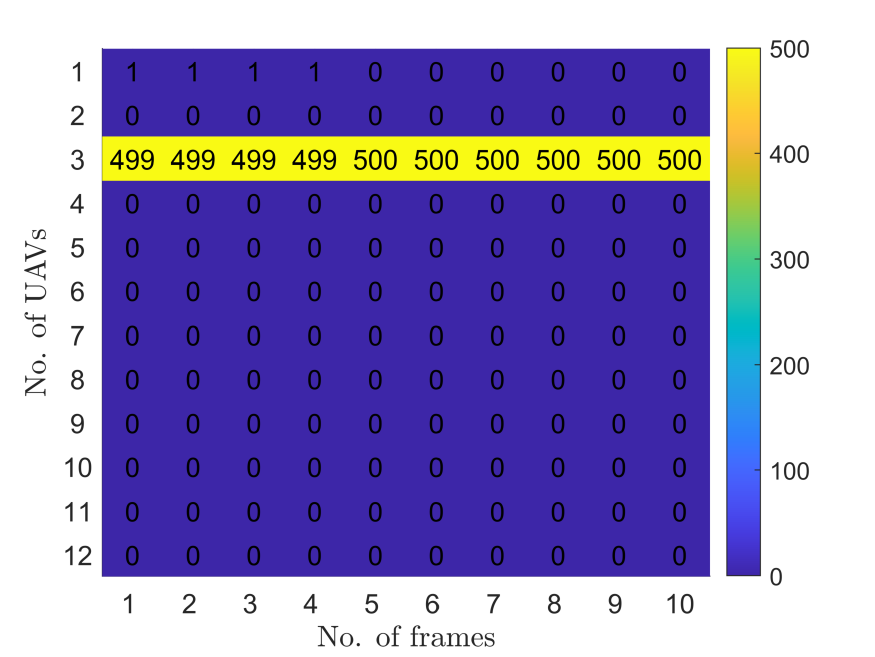}%
				\label{bb}}
			\hfil
			\subfloat[]{\includegraphics[width=2.33in]{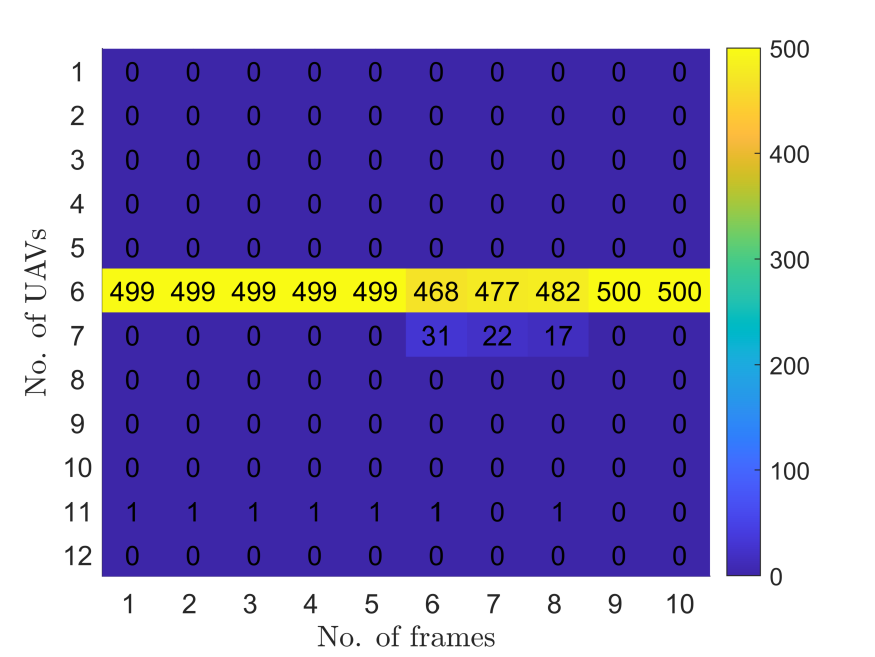}%
				\label{cc}}
			\caption{The bandwidth allocation of UAVs in different frames when $\lambda=0.3$. (a) Bandwidth allocation of UAVs to FAR 1; (b) Bandwidth allocation of UAVs to FAR 2; (c) Bandwidth allocation of UAVs to FAR 3.}
			\label{177}
		\end{figure*}
		
		As mentioned in \textit{Remark 1}, \textcolor{black}{we discuss the performance of the proposed algorithm under different cost factors.
		The weights are set to Setting 1.} The utility under different $\lambda$ in different frames is shown in the Fig. \ref{MDP}. \textcolor{black}{The curves are obtained from the average of more than 100 independent tests.} We find that as $\lambda$ increases, the utility decreases. As can be seen from Eq. \eqref{3}, a larger $\lambda$ means a smaller utility.
		\textcolor{black}{To investigate the impact of the cost factor on the task assignment schemes of UAVs,} the number of UAVs under different $\lambda$ in different frames is shown in the Fig. \ref{1234}. \textcolor{black}{The curves are obtained from the average of more than 100 independent tests.}
		We observe that a larger $\lambda$ means that the number of non-working UAVs increases (i.e., Fig. \ref{1234}(a)), while the number of UAVs suppressing FAR 1, FAR 2, and FAR 3 all decreases (i.e., Fig. \ref{1234}(b), Fig. \ref{1234}(c) and Fig. \ref{1234}(d)). This is because when the cost factor is too large, the cost of UAVs performing suppressive jamming tasks is greater than the reward. To maximize the utility of UAVs, UAVs are more inclined to not work. 
		\textcolor{black}{Thus, the cost factor $\lambda$ could be adjusted flexibly according to the actual combat scenarios. 
		Generally, when the detection capability of FAR network is strong and the stealth effect of UAV is poor, the probability of UAV carrying out jamming mission to be detected increases and it is more likely to be exposed, in this case, the cost factor $\lambda$ could be set larger.}
		
		In addition, the bandwidth allocation of UAVs to different FARs when $\lambda=0$ under an independent trial at different frames are shown in Fig. \ref{777}. From Fig. \ref{777}(a), we find that UAV 10 suppresses FAR 1 at the first frame, and the bandwidth allocation is 500MHz, equal to the working bandwidth of FAR 1. UAV 2 and UAV 10 suppress FAR 1 at the second frame, and the sum of bandwidth allocation is 500MHz, equal to the working bandwidth of FAR 1. 
		Through verification, we find that the sum of bandwidth allocation of UAVs is equal to the FAR 1 working bandwidth in the remaining frames (i.e., frames 3 to frame 10). 
		Similarly, from Fig. \ref{777}(b) and Fig. \ref{777}(c), we also verify that the sum of bandwidth allocation of UAVs is equal to the working bandwidth of FAR 2 and FAR 3, respectively. 
		
		The bandwidth allocation of UAVs to different FARs when $\lambda=0.3$ under an independent trial at different frames are shown in Fig. \ref{177}. By comparing Fig. \ref{777} and Fig. \ref{177}, we find that more UAVs chose not to work, and the bandwidth allocation of working UAVs became larger when $\lambda=0$. This is because, when $\lambda$ is large, there are fewer UAVs working. To ensure full coverage of the FAR working bandwidth, UAVs need to allocate more bandwidth. 
		We also observe that UAV 10 tends to suppress FAR 1, UAV 3 tends to suppress FAR 2 and UAV 6 tends to suppress FAR 3. This is because UAV 10 to FAR 1 is main lobe suppressive jamming, and the antenna gain $G_r^{n,m,k}$ is large, while the remaining UAVs to FAR 1 are side lobe suppressive jamming, and the antenna gain $G_r^{n,m,k}$ is small. Similarly, UAV 3 to FAR 2 is main lobe suppressive jamming, and UAV 6 to FAR 3 is main lobe suppressive jamming.
		
		\subsubsection{Effect of bandwidth estimation error}
		\textcolor{black}{As mentioned in \textit{Remark 2}, we discuss the performance of the proposed algorithm under different bandwidth estimation errors. The inaccuracy of $B_{m}$ estimation will affect the bandwidth allocation of UAVs.} In the simulation above, the $B_{m}$ of each FAR is 500MHz. In the following simulation, assuming that the estimation error of $B_{m}$ obeys a Gaussian distribution with mean 0 and variance $\sigma^2$. The cost factor could be set to $\lambda=0$, and the weights could be set to Setting 1. The utility of proposed algorithm under different $\sigma$ are shown in Table \ref{t511}.
		We observe that a larger $\sigma$ means a larger percentage of decline in utility.
		When the estimated $B_{m}$ is larger than the actual working bandwidth of FAR, the sum of bandwidth allocated by UAVs will also be larger than the actual working bandwidth of FAR, so that there will be a part of excess bandwidth without any suppressive jamming effect.
		When the estimated $B_{m}$ is less than the actual working bandwidth of the FAR, the sum of bandwidth allocated by UAVs will also be less than the working bandwidth of the FAR, so that the sum of bandwidth allocated by UAVs cannot completely cover the working bandwidth of the FAR. 
		If the working frequency of FAR is just within the unsuppressed bandwidth, then FAR will detect high-value target, and the suppressive jamming utility of UAVs is 0.
		\textcolor{black}{In conclusion, inaccurate estimation of $B_{m}$ and large estimation errors $\sigma$ reduce the utility of proposed algorithm.}

		\begin{table*}[t]
			\centering
			\caption{Utility of proposed algorithm under different $\sigma$}
			\label{t511}
			\resizebox{\linewidth}{!}{
				\begin{tabular}{c|c|c|c|c|c|c|c|c|c|c}% 通过添加 | 		
					\toprule
					\hline 
					$k$th frame&$k=1$&$k=2$&$k=3$&$k=4$&$k=5$&$k=6$&$k=7$&$k=8$&$k=9$&$k=10$\\ 
					\hline
					Utility ($\sigma=0$)&0.9478&0.9467&0.9438&0.9393&0.9342&0.9307&0.9275&0.9227&0.9168&0.9070\\
					\hline
					Utility ($\sigma=10\%$)&0.7156&0.7141&0.7122&0.7098&0.7077&0.7074&0.7066&0.7051&0.7027&0.6998\\
					\hline
					Percentage of decline in utility&$\textbf{24.50\%}$&$\textbf{24.58\%}$&$\textbf{24.54\%}$&$\textbf{24.43\%}$&$\textbf{24.25\%}$&$\textbf{24.00\%}$&$\textbf{23.82\%}$&$\textbf{23.59\%}$&$\textbf{23.35\%}$&$\textbf{22.85\%}$\\
					\hline
					Utility ($\sigma=30\%$)&0.6906&0.6886&0.6860&0.6827&0.6796&0.6765&0.6737&0.6693&0.6640&0.6576			\\
					\hline
					Percentage of decline in utility&$\textbf{27.14\%}$&$\textbf{27.26\%}$&$\textbf{27.32\%}$&$\textbf{27.31\%}$&$\textbf{27.25\%}$&$\textbf{27.31\%}$&$\textbf{27.36\%}$&$\textbf{27.46\%}$&$\textbf{27.58\%}$&$\textbf{27.50\%}$\\
					\hline
					Utility ($\sigma=50\%$)&0.6608&0.6584&0.6552&0.6515&0.6481&0.6456&0.6424&0.6385&0.6325&0.6247\\
					\hline
					Percentage of decline in utility&$\textbf{30.28\%}$&$\textbf{30.45\%}$&$\textbf{30.57\%}$&$\textbf{30.64\%}$&$\textbf{30.63\%}$&$\textbf{30.63\%}$&$\textbf{30.74\%}$&$\textbf{30.80\%}$&$\textbf{31.00\%}$&$\textbf{31.12\%}$\\
					\hline
					\bottomrule
				\end{tabular}
			}
		\end{table*}

		\section{Conclusion}
		In this paper, to suppress FAR network effectively, a dynamic UAVs cooperative suppressive jamming method with joint task assignment and bandwidth allocation has been studied. 
		Firstly, the system model constructed has been mathematically characterised as a D-MIP problem. Then, the proposed novel evaluation indicator has evaluated the jamming effect on the FAR network and the devised utility function has quantified the jamming effect of UAVs.
		Finally, the proposed two-step dynamic hybrid algorithm has obtained the allocation schemes with less computational resources in dynamic environment. Extensive numerical simulation results have demonstrated the effectiveness of the proposed algorithm in the aspects of suppressive jamming performance, computational resource saving and dynamic environment adaptability.

		\bibliographystyle{IEEEtran}
		\bibliography{refer}

	\end{document}